\documentclass[11pt,a4paper]{article}
\usepackage{amsmath,amssymb,amsthm,amsfonts}
\usepackage[dvips]{graphicx}
\usepackage{epsfig}
\usepackage{siunitx}	
\usepackage{subfigure}
\usepackage{float}
\usepackage{verbatim}
\usepackage[font=small]{caption}
\usepackage{soul}

\usepackage{color}

\title{Mass transfer from a fluid flowing through a porous media}
\author{T.G. Myers\footnote{Centre de Recerca Matem\`{a}tica, Campus de Bellaterra  Edifici C, 08193 Bellaterra, Barcelona, Spain.}, F. Font\footnotemark[1] }

\def\p{\partial}

\def\({\text{\huge (}}
\def\){\text{\huge )}}

\def\]{\text{\huge ]}}
\def\[{\text{\huge [}}

\providecommand{\keywords}[1]{\textbf{\textit{Keywords:}} #1}

\newcommand{\bi}{\begin{itemize}}
\newcommand{\ei}{\end{itemize}}
\newcommand{\be}{\begin{equation}}
\newcommand{\ee}{\end{equation}}
\newcommand{\ba}{\begin{align}}
\newcommand{\ea}{\end{align}}

\newcommand\nc{\newcommand}
\nc\pad[2]{\frac{\p #1}{\p #2}} \nc\padd[2]{\frac{\p^2 #1}{\p
{#2}^2}} \nc\nd[2]{\frac{d #1}{d #2}} \nc\pat[2]{\frac{D #1}{D
#2}} \nc\ov{\overline} \nc\degree{^{\circ}} \nc\ord[1]{{\cal
O}(#1)} \nc\ra{\rightarrow} \nc\Ra{\Rightarrow} \nc\dint{{\mbox ~
d}}

\newcommand{\bea}{\begin{eqnarray}}
\newcommand{\eea}{\end{eqnarray}}
\newcommand{\beas}{\begin{eqnarray*}}
\newcommand{\eeas}{\end{eqnarray*}}

\begin{document}
\maketitle

\begin{abstract}
A  mathematical model is developed for the process of mass transfer from a fluid flowing through a packed column. Mass loss, whether by absorption or adsorption, may be significant. This is appropriate for example when removing contaminants from flue gases. With small mass loss the model reduces to a simpler form which is appropriate to describe the removal of contaminants/pollutants from liquids. A case study is carried out for the removal of CO2 from a gas mixture passing over activated carbon. Using the experimental parameter values it is shown, via non-dimensionalisation, that certain terms may be neglected from the governing equations, resulting in a form which may be solved analytically using a travelling wave substitution. From this all important quantities throughout the column may be described; concentration of gaseous materials, amount of material available for mass transfer, fluid velocity and pressure. Results are verified by comparison with experimental data for the breakthrough curve (the amount of carbon measured at the column outlet). The advantage of the analytical expression over a purely numerical solution is that it can easily be used to optimise the process. In the final section we demonstrate how the model may be further reduced when small amounts of contaminant are removed. The model is shown to exhibit better agreement than established models when compared to  experimental data for the removal of amoxicillin and congo red dye from water.
\end{abstract}


\section{Introduction}
With oceans overloaded with plastic, the air that we breathe full of noxious substances and even drinking water laced with legal and illegal drugs it is clear that humanity needs to improve its methods for dealing with pollutants. However, cutting down on pollution is not enough, in some cases active removal must be carried out. In this paper we will focus on just one aspect of this issue, namely contaminant removal via column sorption.

Column sorption involves forcing a fluid through a confined tube filled with a porous material capable of removing certain components of the fluid. As the fluid passes through, the component to be removed can attach to the surface of the material (adsorption) or enter into the volume (absorption). This may continue until the material becomes saturated, when no more  removal occurs, and the fluid passes through the material unchanged.
It is perhaps the most popular practical sorption method  \cite{Xu13,Tan12} and is used for a  wide range of processes such as the removal of contaminants including pharmaceuticals, carbon dioxide, heavy metals, dyes and salts \cite{Xu13,Ahmed,Chowd13,Espina,Han,Patel}.

In this paper we will develop a model to describe the flow of a fluid through a porous material contained in a cylindrical column. This configuration has been chosen due to its relevance for a wide number of published experiments and existing extraction equipment. We will focus on a two-component fluid system, with only one component being removed. (The extension to more components is straightforward). Once the model is developed we will analyse it within the context of post-combustion carbon capture and compare with experimental results for breakthrough (that is the CO$_2$ concentration on leaving the column). Subsequently we will show how the model compares with previous, standard models and test it against experimental data for the removal of contaminants such as antibiotics and synthetic dyes.

There exist a wide literature analysing column sorption. A number of simple models focus solely on the column outlet and are based on the probability of a gas molecule escaping \cite{Yoon}. These neglect the evolution of the process through the column and cannot fully explain the physics. However, with carefully chosen parameter values it is possible to  reproduce the breakthrough curves. When flow along the column is accounted for the simplest model balances advection with mass loss, coupled to a rate equation for the mass loss. An early example of this may be found in \cite{Bohart}. Although their full result is often reduced to a much simplified equation valid only at the outlet. Recent models typically deal with variables averaged over the cross-section and include advection, diffusion and temperature effects \cite{Li,BenMansour}. However, since the sorption process is approximated by a simple kinetic model there is still a degree of fitting to match the breakthrough curve. An issue with many numerical studies is highlighted in \cite{Myers19} where it is shown that a number of errors in the governing equations have propagated through the literature. This leads to inaccuracies in the fitting coefficients and therefore incorrect predictions on scaling up.

The study of \cite{Myers19} was based on the assumption of negligible mass removal, so that quantities such as the velocity and fluid density are constant and consequently the pressure gradient is linear.
In the following we will consider a situation where significant quantities are removed, such that the density and velocity may vary along the column. We will follow the style of \cite{Myers19} but do not carry out the averaging which acts to complicate the system. After deriving the governing equations we apply the model to an example of carbon capture, where approximately 15\% of a CO$_2$/N$_2$ mixture is removed by adsorption. Comparison with experimental data for breakthrough shows excellent agreement. Subsequently we compare the full current model, a reduced version of this appropriate for incompressible flow and previous models against data for the removal of amoxicillin and dye from solution.

{The work contains a number of novel elements for this field. The non-dimensionalisation permits us to identify dominant terms and also negligible ones which then permits the  study of a much simplified system (with a known small error). Except during the very initial period, the simplified system permits a travelling wave solution and so we find analytical expressions for all important quantities, concentration, available adsorbent, gas velocity and pressure. These solutions have not previously been published. A full numerical solution is therefore not necessary. It is shown that the gas concentration and available sorbent follow almost identical curves, this means that the experimentally observed breakthrough curve may be used to determine the form of the kinetic relation. }

\section{Derivation of governing equations}\label{GovSec}

Consider a fluid flowing through a porous medium where mass transfer occurs at the fluid-solid interfaces. Within the fluid
the mass continuity equation may be written in the form
\bea
\pad{\rho}{t} + \nabla \cdot \underline{j} = -S \, ,
\eea
where $S$ is a sink term representing mass loss and $\rho$ is the density. The flux is composed of advective and diffusive components
\bea
\underline{j} = -D \nabla \rho + \rho \underline{u} ~ .
\eea
The exact shape of the porous media is unknown hence it is impossible to predict the flow, for this reason it is standard practice to assume plug flow or carry out some radial averaging: the resulting equations are equivalent. Here we will follow the simpler route, assuming plug flow and hence all other variables also only vary with distance along the column.

With the definition $\mathbf{ {u}} = ( {u}(x,t),0)$, mass continuity now specifies
\bea
\pad{\rho}{t} + \pad{( {u}\rho)}{x} = D \padd{\rho }{x} -(1-\epsilon) \rho_{q} M_q   \pad{\overline{q}}{t} 
~ ,
\eea
where $\epsilon$ is the bed void fraction. 
The derivation of the mass loss term requires an assumption that the solid component accepts
material of density $\rho_q$ and molar mass $M_q$  at a rate $\partial \overline{q}/ \partial t$, where $\overline{q}$ represents the amount of material transferred. The overbar indicates it is an average. In practice material is normally adsorbed or absorbed through the surface or in cracks of the solid. In time the surfaces  become saturated. To avoid dealing with an overly complicated system it is standard to ignore the precise distribution of the transferred mass and instead define an average throughout the solid material, which is here denoted $\overline{q}$.
The final term on the right hand side is then a sink representing the mass lost at all solid-fluid interfaces at a given $x$, see \cite{Myers19}.
If the volume flux at the column inlet is $Q_{0}(t)$, then the interstitial velocity may be written $u(x,t)$ with $u(0,t) = {u}_0(t)= Q_{0}(t)/(\epsilon \pi R^2)$. In the literature it is also common to work in terms of the superficial velocity, which is simply the interstitial velocity multiplied by the void fraction, $\epsilon u$.

The density may be expressed in terms of the molar mass $M_i$ and molar concentration $c_i$ of the  components. For a two component system  $\rho = M_1 c_1 + M_2 c_2$, while the molar mass of the gas mixture $M = \rho/(c_1 + c_2)$.

\subsection{Mass transfer}

There are a number of models to approximate the mass transfer process. These are typically based on assumptions that the rate is proportional to the amount available for transfer and the free sites on the solid material and lead to different forms of kinetic model. A classical model is presented in \cite{Bohart}. Local equilibrium, linear driving force, pore diffusion, pseudo-first-order, pseudo-second-order and Avrami models are discussed in \cite{Li,Shaf15}. In the present study we will employ the popular linear kinetic relation
\bea
\label{qlin}
\pad{\overline{q}}{t} & = k_q (\overline{q}^* - \overline{q}) \, ,
\eea
where $k_q$ is a rate constant and $\overline{q}^*$ the saturation value. The form of $\overline{q}^*$ is also the subject of numerous studies, see the review of \cite{Ayawei17}.

The linear kinetic relation has the unrealistic feature that it has a weak dependence on the component available for transfer (which can only enter through the definition of $\bar{q}^*$). It must therefore be specified explicitly that this equation only holds over regions where material is available. There is a tendency in published literature to immediately integrate equation \eqref{qlin} to find $\overline{q} = \overline{q}^* (1-\exp(-k t))$ \cite{Espina,Shaf15,Sarker}.  This is incorrect. Firstly, $\overline{q}^*$ typically depends in a complex way on space and time, secondly the constant of integration is dependent on space. In the special case where $\overline{q}^*$ is constant we may write $\overline{q} = \overline{q}^* (1-\exp(-k (t-t_s(x)))$ where $t_s(x)$ is the time at which the material to be removed first reaches the point $x$. With a typical form of time-dependent $\overline{q}^*$ the integration cannot be carried out.

\subsection{Pressure velocity relations}

If we assume a constant fluid velocity then its value may be easily calculated from the mass or volume flux and the above equations adequately describe the system. This would be appropriate when a negligible mass (compared to the total mass) is removed and is typically the case with transfer from a liquid. If mass removal is significant it will affect  the velocity and pressure. This often occurs in gas mass transfer processes, which can involve the removal of some 20\% of the gas. In this situation we need more equations to close the system. Since the case of gas flow is more complex  henceforth we will focus on that, the liquid case forms a subset of the model developed below.

Firstly, we follow the standard practice of relating the pressure to the concentration via the ideal gas law
\begin{align}
p  =  R_g T   c =  R_g T  ( c_1+c_2)  \, .
\end{align}
The final governing equation required for the system comes from conservation of momentum. For uni-directional plug flow the Navier-Stokes equation may be reduced to the form
\bea
\label{conmom}
\rho\left( \pad{u}{t} + u \pad{u}{x}\right) = -   \pad{ p}{x} + \frac{4}{3}\mu \padd{u}{x} - \textit{S} u \, .
\eea
The term $\textit{S}\, u$  represents a reduction in momentum due to mass loss, where
$\textit{S} =  (1-\epsilon) \rho_{q} M_q \partial\overline{q}/\partial t$ is the  sink from the mass continuity equation.
The 4/3 factor is specific to compressible flow: with a constant density flow this factor becomes unity.

The classical relation describing flow in a porous media is Darcy's law, which was derived for the incompressible flow of a viscous liquid through sand. In the case of a gas flowing through a packed bed Darcy may not be appropriate.
To understand Darcy's law consider conservation of momentum with no mass sink. Assuming the steady, slow flow of an incompressible fluid (where slow is defined such that terms of order $u^2$ may be neglected) equation \eqref{conmom} reduces to
\bea
0 = -  \pad{ p}{x} + \mu \padd{u}{x} \, .
\eea
The key assumption is that viscous resistance is proportional to  velocity $u_{xx} \propto u$ and then we obtain
Darcy's law
\bea
-\pad{p}{x} = \frac{\mu }{k_p} u \, ,
\eea
where $k_p$ is termed the permeability.
Gases have a very low viscosity so it is possible that for gas flow viscous resistance is small and instead inertial resistance plays a significant role. When inertia is non-negligible a drag law is usually invoked
\bea
u u_x  =   C_D \frac{u^2}{d}
\eea
where the length-scale $d$ is chosen as the typical particle diameter and the drag coefficient $C_D=C_D(\epsilon, Re, d)$. In a similar manner to the derivation of Darcy's law we may use this expression to obtain a relation between pressure gradient and velocity.

For the current problem, the momentum equation also contains a sink term proportional to velocity. Including both viscous and inertial resistance as well as the sink we  obtain the pressure-velocity relation
\bea
-\pad {p}{ x} =  \alpha  \rho u^2 + \left(\beta + S\right) u \, ,
\label{pxu}
\eea
where $\alpha, \beta$ are constant while  $S$ is variable.
In the absence of the sink term this form of equation is discussed in \cite{Shaf14}. Various values for the constants $\alpha, \beta$ reproduce the Ergun relation (applicable to beds packed with beads or granules), or forms appropriate to beds packed with cylinders, foams and laminates \cite{Reza09}. With $\alpha=S=0$ Darcy's law is retrieved. Including  the sink term we have not been able to find this relation in the sorption literature, but it is a natural consequence of  momentum conservation with mass loss.
Away from the sorption front we expect the sink term to be small and its neglect is justifiable, however in regions of rapid mass transfer it plays a significant role in the mass balance and it is possible that this also leads to significant momentum loss.

\subsection{Summary of governing equations}

Noting that
$ \rho = M_1 c_1 + M_2 c_2$ we may  eliminate the gas density from the mass balance
\begin{align}
\pad{ }{t}(M_1 c_1 + M_2 c_2)  + \pad{}{x}(u(M_1 c_1 + M_2 c_2))&= D  \padd{ }{x}(M_1 c_1 + M_2 c_2) -  (1-\epsilon) \rho_{q} M_q   \pad{\overline{q}}{t}  \, .
\end{align}
Conservation of each species then gives
\begin{align}
\pad{ c_1}{t}  + \pad{}{x}(u c_1  ) &= D  \padd{ c_1 }{x}  -  (1-\epsilon)  \rho_{q} \frac{M_q}{M_1}   \pad{\overline{q}}{t}   ~ ,\label{c1eq} \\
\pad{c_2 }{t}  + \pad{}{x}(u c_2) &= D  \padd{ c_2}{x} \, \label{c2eq}.
\end{align}
If only one species is removed we may set $M_q = M_1$ that is, the molar mass of the component is the same in the solid and fluid state.
The remaining governing equations are
\begin{align}
\pad{\overline{q}}{t} & = k_q (\overline{q}^* - \overline{q})\, \label{qeq}\\
p  & =    R_g T  ( c_1 +   c_2) \, ,\label{peq}\\
-\pad {p}{ x} &= \alpha (M_1 c_1 + M_2 c_2) u^2 + \left(\beta +(1-\epsilon) \rho_{q} M_1   \pad{\overline{q}}{t}\right) u
~ . \label{pxeq}
\end{align}
Equations (\ref{c1eq} -- \ref{pxeq}) provide the system to describe the evolution of the five variables $c_1,  c_2, u, p, \overline{q}$. The five equations are required when mass transfer is significant compared to the total mass flow.
When dealing with the transfer of trace amounts of a material, for example drugs in the water supply, the system is significantly simpler. First, the velocity is approximately constant $u = Q_0/(\epsilon \pi R^2)$ and the problem is described by equations (\ref{c1eq}, \ref{qeq}). If a known pressure drop drives the flow then the velocity may also be calculated by neglecting the $c_1$ and $\overline{q}$ terms in equation \eqref{pxeq} (since both terms are small when trace amounts are removed).

{The assumptions made in deriving the above equations include: the amount of material transferred may be averaged over the sorbent; $\overline{q}_t$  follows a linear kinetic model (this is discussed later); an ideal gas law holds; the flow may be approximated as plug flow.}

\subsection{Boundary and initial conditions}

At the inlet, $x=0$, there is continuity of mass flux so
\bea
\left. (\rho u)\right|_{x=0^-}  = \left. \left( \rho u - D \pad{ {\rho}}{x}\right)\right|_{x=0^+} \, ,
\eea
where the $-, +$ superscripts indicate just before and just after $x=0$.
Separating the concentrations gives
\begin{align}
u(0^-,t) c_{10}    &=   \left. \left(uc_1 - D \pad{c_1}{x}\right)\right|_{x=0^+} \, , \quad
u(0^-,t) c_{20}    =  \left. \left( uc_2 - D \pad{c_2}{x}\right)\right|_{x=0^+}  \, ,
\end{align}
where
$c_{10}$,  $c_{20}$ are the concentrations of the inlet gas and
$u(0^-,t) = Q_0/(\pi R^2)$, $u(0^+,t)=Q_0/(\epsilon \pi R^2)$.

At the exit we move to a region where no mass transfer occurs, so it is assumed that whatever the density on leaving the column it remains the same just outside the exit 
\bea
\left.   \pad{c_1}{x} \right|_{x=L^-} = \left.   \pad{c_2}{x} \right|_{x=L^-} = 0 \, .
\eea
The pressure at inlet and outlet are
\bea
{p}(0,t)= p_0(t) \, , \qquad  {p}(L) = p_a \, .
\eea
The inlet value may vary with time.
Initially we assume the solid is fresh and the column is free of the component to be removed
\bea
\rho(x,0) = M_2 c_2(x,0) = M_2 \frac{p(x)}{RT} ~, \qquad q(x,0)=0 ~ ,
\eea
hence
\bea
c_1(x,0) = 0 ~, \qquad  c_2(x,0) =  \frac{p_{in}(x)}{RT} \, ,
\eea
where the initial pressure $p_{in}(x)=p_0(0) - (p_0(0)-p_a)x/L$.

\section{Non-dimensionalisation}
We scale variables in the following manner
\begin{align}\label{nondim}
\hat{p} & = \frac{p-p_a}{\Delta p} \quad \hat{c}_1 = \frac{c_1}{c_{10}} \quad \hat{c}_2 = \frac{c_2}{c_{20}}\quad \hat{q} = \frac{\overline{q}}{\overline{q}_0^*} \quad
\hat{x} =\frac{x}{{\cal L}} \quad \hat{t} = \frac{t}{\Delta t} \quad \hat{u} =\frac{u}{u_0}\, ,
\end{align}
where ${\cal L}$ and $\Delta t$ are unknown and $\overline{q}_0^*$ is the value of  $\overline{q}^*$ at  $t=0$.
Since our interest lies with the reaction we choose a time-scale $\Delta t = 1/k_q$ and so
\bea
\pad{\hat{q}}{\hat{t}} & = (\hat{q}^* - \hat{q}) \, \label{NDqeq} .
\eea
Balancing advection with mass loss gives the length-scale ${\cal L} = u_0 c_{10}/((1-\epsilon) \rho_q \overline{q}_0^* k_q)$ and then
\begin{align}
\delta_1 \pad{ \hat{c}_1}{\hat{t}}  + \pad{}{\hat{x}}(\hat{u}  \hat{c}_1  ) &= \delta_2  \padd{ \hat{c}_1 }{\hat{x}}  -  \pad{\hat{q}}{\hat{t}}   ~ ,\label{NDc1eq} \\
\delta_1 \pad{\hat{c}_2 }{\hat{t}}  + \pad{}{\hat{x}}(\hat{u}    \hat{c}_2) &= \delta_2  \padd{ \hat{c}_2}{\hat{x}} \, \label{NDc2eq}.
\end{align}
In  experiments the main gas component is not usually the one being removed, hence we assume $c_{10} < c_{20}$ and write
\begin{align}
1+ \delta_3  \hat{p} & =  \delta_4  \left( \hat{c}_2 + \delta_5 \hat{c}_1 \right) \, ,\label{NDpeq}\\
-\pad{\hat{p}}{\hat{x}} &= \delta_6 (\hat{c}_2 + \delta_7 \hat{c}_1 ) \hat{u}^2 +   \left(1  + \delta_8  \pad{\hat{q}}{\hat{t}}\right) \hat{u} \,\label{NDpxeq} ,
\end{align}
where, assuming the flow to be close to Darcy flow, we have set the pressure scale $\Delta p = \beta u_0 {\cal L}$.

The boundary and initial conditions are
\begin{align}
1    &=   \left. \left(\hat{u}\hat{c}_1 - \delta_2 \pad{\hat{c}_1}{\hat{x}}\right)\right|_{\hat{x}=0^+} \, , \quad
1    =   \left. \left(\hat{u}\hat{c}_2 - \delta_2 \pad{\hat{c}_2}{\hat{x}}\right)\right|_{\hat{x}=0^+} \, , \\
\left.   \pad{\hat{c}_1}{\hat{x}} \right|_{\hat{x}=\hat{L}^-} & = \left.   \pad{\hat{c}_2}{\hat{x}} \right|_{\hat{x}=\hat{L}^-} = 0 ~ , \qquad \hat{p}(0,\hat{t})=\hat{p}_0(\hat{t}) ~, \quad \hat{p}(\hat{L})=0 ~ ,\\
\hat{c}_1(\hat{x},0) &= 0  ~ , \qquad \delta_4 \hat{c}_2(\hat{x},0) = 1+ \delta_3 \hat{p}_{in}  ~ , \qquad  {\hat{q}}(\hat{x},0) = 0 ~ , \label{ic_non}
\end{align}
where the initial pressure profile $\hat{p}_{in}= \hat{p}_0(0) (1- \hat{x}/\hat{L})$. To keep the model general we express the pressure condition at $\hat{x}=0$ as an unspecified function of time. If the inlet pressure is kept constant at a given value then $\hat{p}(0,\hat{t})=1$, however in a  number of experiments the flux is maintained as the constant (via a flow meter) and this requires a variable inlet pressure. This is then determined as part of the solution process.

The five equations (\ref{NDqeq} -- \ref{NDpxeq}), together with the appropriate boundary conditions, are sufficient to determine the five unknowns $\hat{c}_1, \hat{c}_2, \hat{q}, \hat{p}, \hat{u}$. They contain eight non-dimensional groupings, $\delta_i$, which indicate the relative importance of the terms
\begin{align}
\delta_1 &= \frac{{\cal L} k_q}{u_0} = \frac{  c_{10}}{(1-\epsilon) \rho_q \overline{q}_0^*}\, , \quad  \delta_2 = \frac{D}{{\cal L} {u}_0} \, ,\quad \delta_3 = \frac{\Delta p}{p_a} \, ,\quad
\delta_4 = \frac{ R_g T   c_{20}}{p_a}
\label{d1tod4}
\\
\delta_5  & = \frac{ c_{10}}{ c_{20}}\, ,\quad
\delta_6  = \frac{\alpha M_2 c_{20} u_0^2 {\cal L}}{\Delta p} \, ,\quad
\delta_7  = \frac{M_1 c_{10}}{M_2  c_{20}} \, ,\quad \delta_8 =
\frac{1-\epsilon}{\beta } \rho_{q} M_1 \overline{q}_0^* k_q \, .\label{d5tod8}
\end{align}

\section{Application to carbon capture in a packed column}\label{CCSec}

Standard models for carbon capture in a packed column may be found in the reviews of \cite{Li,BenMansour,Shaf14}. In general these are based on the following assumptions:
\begin{enumerate}
\item The gas behaves as an ideal gas.
\item A plug-flow model is adopted.
\item  The radial variation of concentration is negligible.
\item The bed operates isothermally.
\item The CO$_2$ concentration is low so that the pressure gradient is linear and the velocity constant
along the bed.
\end{enumerate}
Assumption 3 follows from 2 (although in mathematical terms the radial variation is the source of the mass sink term, see \cite{Myers19}). In experimental studies CO$_2$ typically comprises 15-20\% which, as we will see later, leads to velocity variation of the same order and non-linear pressure, hence we do not apply assumption 5. We have retained the isothermal and ideal gas assumptions (in \cite{Myers19} it was shown that the temperature variation is small during carbon capture).

A typical experimental set-up involves a circular cross-section column containing a porous material, this is then placed inside an oven or furnace to regulate the temperature. Gas is passed through the column and the concentration measured at the outlet. Here we consider a specific experiment which involves a CO$_2$, N$_2$ mixture passing through a bed of activated carbon, the data and operating conditions are given in Table \ref{tab:Table1}, see  \cite{Myers19,Shaf15}. Mass transfer was by adsorption. A mass flow controller was employed to maintain a constant flow rate $Q_0 = 50$ml/min, which indicates $u_0 = 50 \times 10^{-6}/(60 \epsilon \pi R^2) = 0.019$m/s. A full description of the experiment may be found in \cite{Shaf15}.

\begin{table}
\centering
\begin{tabular}{cccc}
\hline
& Symbol & Value & Dimension \\
\hline
Initial concentration (CO$_2$)  & $c_{10}$ & 6.03 & mol/m$^3$ \\
Initial concentration (N$_2$)  & $c_{20}$ & 34.19 & mol/m$^3$ \\
Molar mass (CO$_2$)  & $M_{1}$ & 0.044 & kg/mol \\
Molar mass (N$_2$)  & $M_{2}$ & 0.028 & kg/mol \\
Temperature & $T$ & 303.15 & K \\
Ambient pressure & $p_a$ & $101325 (1) $ & Pa (Atm) \\
Adsorption saturation   & $\bar{q}^*$ & 1.57 & mol/kg \\
Bed void fraction & $\epsilon$ & 0.56 & - \\
Bed length & $L$ & 0.2 & m \\
Bed radius & $R$ & 0.005 & m \\
Diameter of bed particles & $d_p$ & 6.5 $\times 10^{-4}$ & m \\
Gas viscosity (15\% CO$_2$ (1.5), 85\% N$_2$ (1.8)) & $\mu_g$ & $1.76 \times 10^{-5}$ & Pa s \\
Density  of adsorbed CO$_2$ & $\rho_{q}$ & 325 & kg/m$^3$ \\
Axial diffusion coefficient & $D$ & $2.57 \times 10^{-5}$ & m$^2$/s \\
Initial volume fraction (CO$_2$) & $y_1$ & 0.15 & - \\
Initial volume flux & $Q$ & $8.3\times 10^{-7}$ & m$^3$/s \\
Initial interstitial velocity  & $u_0$ & 0.019 & m/s \\
Adsorption rate constant (CO$_2$) & $k_q$ & 0.0137 & s$^{-1}$ \\
Solid/gas density & $\rho_s/\rho_g$ & 1818/1.2 & kg/m$^3$\\
\hline
\end{tabular}
\caption{Values of the thermophysical parameters mainly taken from \cite{Shaf15}, except $k_q, \rho_{q}$ (as discussed later).}
\label{tab:Table1}
\end{table}

We assume that the flow rate within the porous media is described by the Ergun relation, hence the constants of equation \eqref{pxeq} are defined by
\bea
\alpha = \frac{1.75 (1-\epsilon)}{d_p \epsilon} \approx 2.11 \times 10^3 \qquad \beta = \frac{150 \mu_g (1-\epsilon)^2}{d_p^2 \epsilon^2} \approx 3.86 \times 10^3 \, ,
\eea
where $d_p = 6.5 \times 10^{-4}$m.
The length-scale ${\cal L} \approx 3.75$cm which indicates the size of the reaction zone. The pressure scale $ \Delta p = \beta u_0 L \approx 14.62$Pa, then using values from Table \ref{tab:Table1} and the definitions (\ref{d1tod4},\ref{d5tod8})
we determine
\bea
\delta_1 = 0.027 \quad \delta_2 \sim 0.036 \quad \delta_3 \sim 1.44 \times 10^{-4} \quad \delta_4 = 0.85 \quad \delta_5 \approx 0.18 \\ \delta_6 = 3.3\times 10^{-3} \quad \delta_7 = 0.28 \quad \delta_8  = 3.47 \times 10^{-5} \, .
\eea

For a kinetic model we take the version described in \cite{Shaf15} since this provides the  parameter values for this specific experiment,
\bea
\overline{q}^* = \frac{q_{m1} K_{T1} p}{[1+(K_{T1} p)^{n1}]^{1/{n1}}}+\frac{q_{m2} K_{T2} p}{[1+(K_{T2} p)^{n2}]^{1/{n2}}} ~ ,
\eea
where $q_{m1}, q_{m2} = 0.69, 3.57$ mol/kg, $K_{T1}, K_{T2}= 8.14\times 10^4, 0.66$ atm$^{-1}$, $n1, n2=0.27, 0.65$ Note, the variation with temperature discussed in that paper does not apply to the present isothermal study. In non-dimensional form we write
\bea
\overline{q}_0^*\hat{q}^* = \frac{q_{m1} K_{T1} p_a (1+\delta_3 \hat{p})}{[1+(K_{T1} p_a^{n1} (1+\delta_3 \hat{p}))^{n1}]^{1/{n1}}}+\frac{q_{m2} K_{T2} p_a (1+\delta_3 \hat{p})}{[1+(K_{T2} p_a^{n2} (1+\delta_3 \hat{p}))^{n2}]^{1/{n2}}} \, .
\eea
Neglecting terms of order $\delta_3 \sim 10^{-4}$ we see that the leading order
equation for $\hat{q}$ is
\bea
\pad{\hat{q}}{t} = 1 - \hat{q} \label{LOqeq}
\eea
where the (constant) adsorbent scaling
\bea
\overline{q}_0^* = \frac{q_{m1} K_{T1} p_a  }{[1+(K_{T1} p_a^{n1}]^{1/{n1}}}+\frac{q_{m2} K_{T2} p_a }{[1+(K_{T2} p_a^{n2}]^{1/{n2}}} \, .
\eea
Neglecting terms of order $\delta_1, \delta_2 \sim 10^{-2}$ the
concentration equations are
\begin{align}
\pad{}{\hat{x}}(\hat{u}  \hat{c}_1  ) &=   -  \pad{\hat{q}}{\hat{t}}   ~ ,\label{LOc1eq} \\
\pad{}{\hat{x}}(\hat{u }   \hat{c}_2) &= 0 \, \label{LOc2eq}.
\end{align}
These simply state that gas is primarily advected through the column, with the only variation coming through the mass loss due to adsorption. In a constant velocity study \cite{Myers19} the leading order $c_1$ equation was discussed, where it was pointed out that although diffusion is, in general small, it can play an important role in the numerical solution. If we have an initial condition where the CO$_2$ concentration is zero and then jumps to some non-zero value when first pumped in then the gradient is discontinuous and diffusion is important in smoothing this out. However, this is only important near $x=t=0$. For the present we will focus mainly on the outlet, close to first breakthrough and so neglect diffusion (except in Fig. \ref{NumTW} where we compare the approximate solutions with numerics to verify the validity of this approach).

The leading order pressure relations are
\begin{align}
1  & =  \delta_4  \left( \hat{c}_2 + \delta_5 \hat{c}_1 \right) \, ,\label{LOpeq}\\
-\pad{\hat{p}}{\hat{x}} &=   \hat{u} \,\label{LOpxeq} .
\end{align}
The form of the non-dimensional gas law, \eqref{LOpeq}, is very informative: the relative change in pressure along the column is tiny compared to the total pressure, hence the pressure variation along the column has a negligible effect on the concentration and we may write one concentration in terms of the
other, irrespective of the pressure or velocity,
\bea
\hat{c}_2 =  \frac{1}{\delta_4} - \delta_5 \hat{c}_1 \, .\label{c2c1eq}
\eea
The scaling of the momentum balance \eqref{LOpxeq} shows that for this case the Ergun relation is unnecessarily complicated: momentum loss is dominated by viscous resistance between the gas and porous material. Hence velocity is related to pressure drop via a simple Darcy law.

The necessary leading order boundary and initial conditions are
\begin{align}
1    &=   \left. \left(\hat{u}\hat{c}_1 \right)\right|_{\hat{x}=0^+} \, \qquad
1    =   \left. \left(\hat{u}\hat{c}_2 \right)\right|_{\hat{x}=0^+} \, , \\
\left.   \pad{\hat{c}_1}{\hat{x}} \right|_{\hat{x}=\hat{L}^-} & = \left.   \pad{\hat{c}_2}{\hat{x}} \right|_{\hat{x}=\hat{L}^-} = 0
\qquad  {\hat{q}}(\hat{x},0) = 0 \, .
\end{align}
Integration of equation \eqref{LOc2eq} shows that $\hat{u} \hat{c}_2 = 1$ everywhere.  Combining this with equation \eqref{c2c1eq} transforms the mass balance \eqref{LOc1eq} to
\begin{align}
\delta_4 \pad{}{\hat{x}}\left(  \frac{\hat{c}_1}{1-\delta_{45} \hat{c}_1}  \right) &=   -  \pad{\hat{q}}{\hat{t}}   ~  ,\label{LOc1}
\end{align}
where $\delta_{45} = \delta_4 \delta_5 \approx 0.15$. The factor $\delta_4/(1-\delta_{45} \hat{c}_1)$ does not appear in the constant velocity system of \cite{Myers19}. Since $\delta_4 = 0.85$ this indicates a change in size of quantities of the order 15\% when compared to constant velocity models. Equation \eqref{LOc1} must be solved in conjunction with the adsorbent equation, \eqref{LOqeq} and then $\hat{u}, \hat{c}_2, \hat{p}$ may be obtained from the preceding equations.

The above equations hold behind the reaction front, $\hat{x} \le \hat{s}$. Ahead we have $\hat{c}_1 = \hat{q} = 0$ and so, from \eqref{c2c1eq}, $\hat{c}_2 = 1/\delta_4$ is constant and hence so is $\hat{u} = \delta_4$ (i.e. the velocity ahead of the front is 15\% lower than the inlet velocity).

\subsection{Travelling wave}
Equations (\ref{LOqeq}, \ref{LOc1}) may be solved numerically to find the behaviour for all time although, due to the neglect of diffusion, the very early time solution may be inaccurate. For sufficiently large times, such that the initial transient is complete, we do not even need a numerical solution since there exists a travelling wave solution. To find this we first
choose a co-ordinate moving with the $c_1$ front, $\eta = \hat{x}-\hat{s}(\hat{t})$. Equations (\ref{LOqeq}, \ref{LOc1}) then become
\begin{align}
\pad{\hat{f}}{\hat{\eta}}  &=   \hat{s}_{\hat{t}}  \pad{\hat{q}}{\eta}   ~  ,\label{feta}\\
-\hat{s}_{\hat{t}}  \pad{\hat{q}}{\hat{\eta}} & = (1 - \hat{q}) ~ ,\label{qeta}
\end{align}
where $\hat{f} = \hat{u} \hat{c}_1 = \delta_4 \hat{c}_1/(1-\delta_{45} \hat{c}_1)$. A travelling wave solution may be found if $\hat{s}_{\hat{t}} = \hat{v}$ is constant. If this is the case the equation for $\hat{f}$ may be integrated immediately. After applying the boundary conditions $\hat{c}_1 = \hat{f}=\hat{q}= 0$ at $\hat{\eta} = 0$ we obtain
\begin{align}
\hat{f} &=   \hat{v}   \hat{q}    ~ .\label{fq}
\end{align}
Eliminating $\hat{q}$ between (\ref{feta} -- \ref{fq}) leads to an ODE
\bea
\hat{v} \pad{\hat{f}}{\hat{\eta}} - \hat{f} = -\hat{v}
\eea
with solution
\bea
\label{fveq}
\hat{f} = \hat{v}\left[1-e^{\hat{\eta}/\hat{v}}\right] \, .
\eea
The constant of integration in this case has been obtained by applying the condition at $\hat{\eta} = 0$. The velocity may be determined by
assuming the front is far from the column entrance, $\hat{x}=0$, hence $\hat{\eta} = \hat{x}-\hat{s}$ is  large and negative, formally we apply $\hat{\eta} \rightarrow -\infty$, $\hat{u} \hat{c}_1 = \hat{f}  \rightarrow 1$ which  determines $\hat{v} = 1$. This verifies the travelling wave assumption that $\hat{s}_{\hat{t}}$ is constant.

The CO$_2$ concentration may be calculated from \eqref{fveq}, with $\hat{v}=1$,
\bea
\label{c1etaeq}
\hat{c}_1 = \frac{1-e^{\hat{\eta}}}{\delta_4+\delta_{45}(1- e^{\hat{\eta}})} \approx \frac{1-e^{\hat{\eta}}}{1-\delta_{45} e^{\hat{\eta}}}
\, .
\eea
The simplification results from $\delta_4+\delta_{45} = {p}_0/{p}_a=1+\delta_3 \approx 1$. Then from (\ref{c2c1eq}, \ref{fq}) and using  $\hat{u} \hat{c}_2 = 1$, $\delta_4 \approx 1- \delta_{45}$ we find
\bea
\hat{c}_2 = \frac{1}{1-\delta_{45} e^{\hat{\eta}}} ~ ,\qquad \hat{u} = 1-\delta_{45} e^{\hat{\eta}} ~, \qquad  q = 1-e^{\hat{\eta}}\, .
\label{c2uqeq}
\eea

The pressure equation \eqref{LOpxeq} may be written as $\hat{p}_{\eta} = - \hat{u}$. For $\hat{x }\ge \hat{s}$, $\hat{u} = \delta_4$ is constant and with $\hat{p}=0$ at $\hat{\eta}=\hat{L}-\hat{s}$ we obtain
\bea
\hat{p} = \delta_4 ((\hat{L}-\hat{s})- \hat{\eta })  \qquad \mbox{for} ~ \hat{x} \ge \hat{s} ~ .\label{p57}
\eea
For $\hat{x }\le \hat{s}$ the velocity is specified by \eqref{c2uqeq}. The pressure at the column inlet is unknown but equation \eqref{p57} gives the value at $\hat{x}=\hat{s}$ ($\hat{\eta}=0$), which may be used to determine the constant of integration, leading to
\bea
\hat{p} = \delta_4  (\hat{L}-\hat{s})-\left[\hat{\eta} + \delta_{45}(1-e^{\hat{\eta}})\right] \qquad \mbox{for} ~ \hat{x} \le \hat{s} ~ .
\eea

To relate the solutions to experiments requires switching between $\hat{\eta}$ and $\hat{x}, \hat{t}$. With the definition $\hat{s}_{\hat{t}}  = 1$ we obtain $\hat{s} = \hat{t} +\hat{s}_0$ (where $\hat{s}_0$ is unknown) and hence $\hat{\eta} = \hat{x} - \hat{t} -\hat{s}_0$. The travelling wave is not valid at $\hat{t}=0$ so we cannot apply an initial condition to determine $\hat{s}_0$ and must use information from a later time. In the carbon capture literature the breakthrough curve is generally presented, which shows the  CO$_2$ concentration at the end of the column. If  CO$_2$ is first recorded at the outlet at time $\hat{t}_b$, then we may use $\hat{s}_0 = \hat{L}- \hat{t}_b$. However, given the uncertainty as to when gas actually first escapes, a more reliable measure is to note the time $t_{1/2}$ when $c_1=c_{10}/2$ and then solve equation \eqref{c1etaeq} with $\hat{x}=\hat{L}$, $\hat{c}_1=1/2$ to determine $\hat{s}_0 = \hat{L} - \hat{t}_{1/2} + \log (2-\delta_{45})$ or alternatively we may write $\hat{t}_b=\hat{t}_{1/2} - \log (2-\delta_{45})$.

\subsection{Dimensional solutions}

We now summarise, in dimensional form, the solutions starting with the simplest region, ahead of the front $x \ge s$, where $s = {\cal L} k_q t + s_0 = {L}+{\cal L} k_q(t-{t}_b)$, and $t_b$ could be set as the experimentally measured breakthrough time or we use $t_b = t_{1/2} - \log(2-( R_g T c_{10}/p_a))/k_q$ and the length-scale ${\cal L} =  u_0 c_{10}/((1-\epsilon) \rho_q \overline{q}_0^* k_q)$. The velocity of the front $\hat{s}_{\hat{t}}=1$ indicates $s_t = {\cal L}/\Delta t = u_0 c_{10}/((1-\epsilon) \rho_q \overline{q}_0^*)$.

For $ x \in [s, L]$ and $t < t_b$:
\begin{align}
c_1 &= 0 ~, \quad c_2 = \frac{p_a}{ R_g T} ~, \quad \bar{q} = 0 ~, \quad u = \frac{ R_g T c_{20}}{p_a} u_0 \label{c1c2quDim}\\
p &= p_a  +\beta \frac{R_g T}{p_a} u_0 c_{20} (L - x) ~ .\label{pdimxgs}
\end{align}

For $x \in [0,s]$
\begin{align}
c_1 &= c_{10} \left[ \frac{1-e^{(x-L)/ {\cal L}}e^{-k_q (t-t_b)}}{1-( R_g T c_{10}/p_a) e^{(x-L)/\hat{v} {\cal L}}e^{-k_q (t-t_b)}} \right]  ~ , \label{c1dim}\\
c_2 &= c_{20} \left[ \frac{1}{ 1-( R_g T c_{10}/p_a) e^{(x-L)/{\cal L}}e^{-k_q (t-t_b)}} \right] ~ ,
\\
\bar{q} &= \bar{q}_0^* \left(1-e^{(x-L)/ {\cal L}}e^{-k_q (t-t_b)}\right) \label{qdim} ~ ,\\
u &=  u_0 \left(1-( R_g T c_{10}/p_a) e^{(x-L)/{\cal L}}e^{-k_q (t-t_b)}\right)  ~ ,\label{udim}\\
\label{pdimxls}
p &= p_a + \beta u_0 \left[ (s-x) -
\frac{ R_g T}{p_a} c_{10}  {\cal L} (1-e^{(x-L)/ {\cal L}}e^{-k_q (t-t_b)}) +
\frac{ R_g T}{p_a}  c_{20}  (L-s) \right] ~ .
\end{align}

In experiments it is common to measure
the breakthrough curve. The analytical representation is obtained by setting $x=L$ in equation \eqref{c1dim}
\begin{align}
c_1 &= c_{10} \left[ \frac{1- e^{-k_q (t-t_b)}}{1-( R_g T c_{10}/p_a) e^{-k_q (t-t_b)}} \right]   ~ ,\label{c1Break}
\end{align}
where $t \ge t_b$.

\subsection{Fixed pressure solution}

In the case studied above the flux is fixed by a flowmeter: the pressure is  adjusted to maintain a constant flow rate.  However it is also common practice to fix the pressure at either end and leave these constant throughout the experiment. For a fixed flux (constant $u_0$) equations (\ref{pdimxgs}, \ref{pdimxls}) determine the pressure ahead of and behind the front respectively for the specified value of $u_0$. If the pressure is fixed such that $p(0,t)=p_0$, $p(L,t)=p_a$ then ahead of the front equation \eqref{pdimxgs} holds, while behind
\begin{align}
p &= p_0 - \beta u_0 \left[ x -
\frac{R_g T}{p_a} c_{10} {\cal L} e^{-k_q (t-t_b)} \left(e^{(x-L)/ {\cal L}}
-e^{-L/{\cal L}}\right) \right] \quad & x \le s ~ .
\end{align}
Since the pressure is continuous at the front we may equate these two expressions to determine the velocity $u_0$ caused by the prescribed pressure drop
\bea
u_0 = \frac{p_0-p_a}{\beta} \left[s + \frac{  R_g T}{{p}_a}\left( c_{20}(L-s)  -
{c}_{10}  {\cal L} e^{-k_q (t-t_b)} \left(e^{(s-L)/{\cal L}}
-e^{-L/{\cal L}}\right) \right)\right]^{-1}
~ .
\eea
The inlet flux $Q_0(t) = \epsilon \pi R^2 u_0(0,t)$ is then obviously time-dependent.

\subsection{Carbon capture results}

\begin{figure}
\centering
\includegraphics[width=0.7\textwidth]{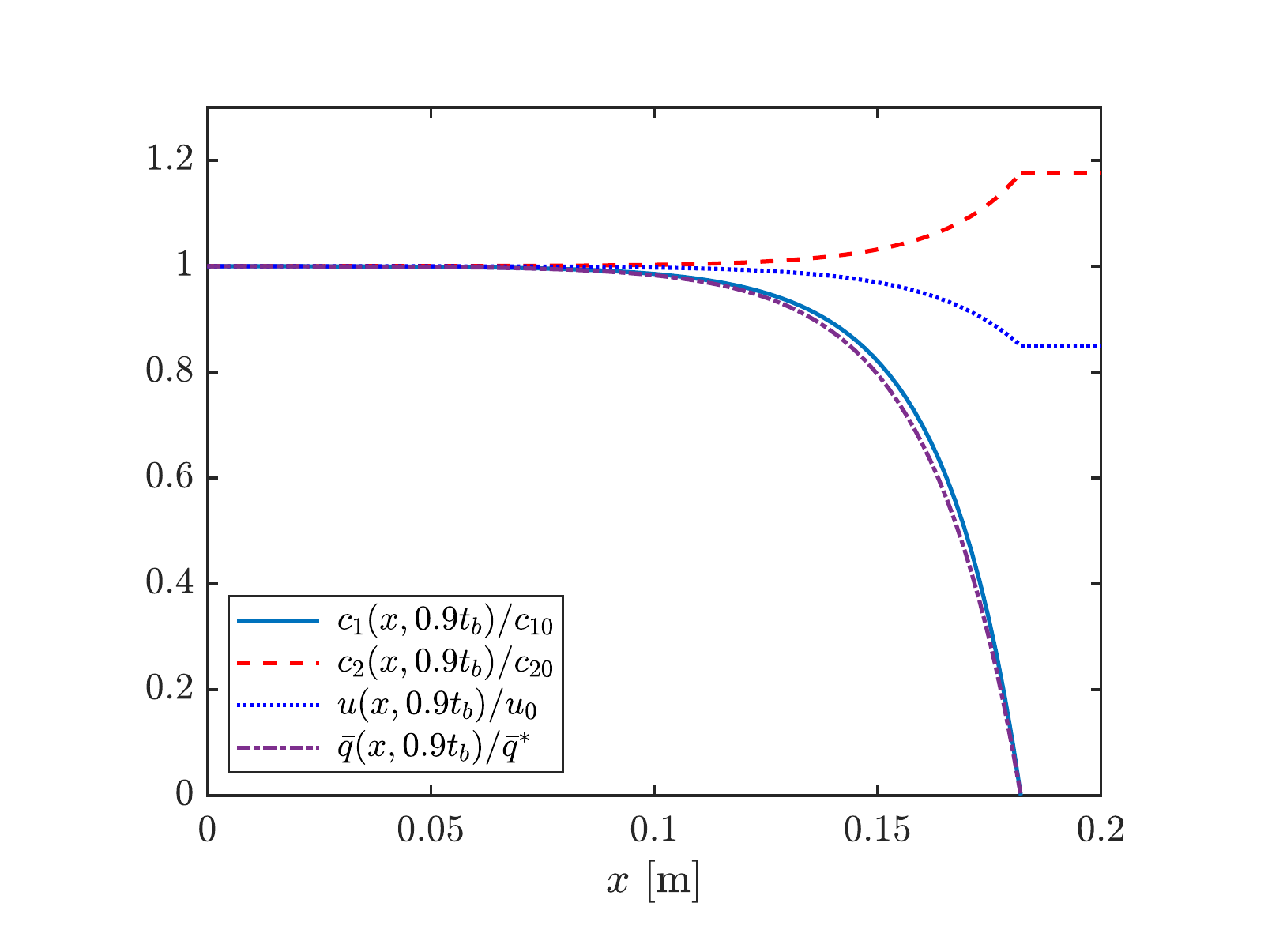}
\caption{Concentrations $ {c}_1(x,t),  {c}_2(x,t)$, amount adsorbed $\bar{q}(x,t)$ and velocity $ {u}$ {along the column} at $t=0.9t_b$}
\label{c12u}
\end{figure}

In Figure \ref{c12u} we show  the variation along the column of the concentration of CO$_2$, N$_2$, the amount of adsorbed material and the velocity of the mixture as predicted by equations (\ref{c1c2quDim}, \ref{c1dim}-\ref{udim}). The parameter values  used are provided in Table \ref{tab:Table1}. The results correspond to a time $t=0.9 t_b$, where $t_b = 10.9$ min is the breakthrough time quoted in \cite{Shaf15}.  As expected from a physical point of view, the CO$_2$ concentration is almost unity at the inlet and decreases smoothly to zero at the front.  The non-dimensionalisation shows that $\hat{q}$ differs from $\hat{c}_1$ by a factor of order $1/(1-\delta_{45})$, where $\delta_{45} \approx 0.1$, and so $c_1$ must  be very similar to $\bar{q}$. This may be observed through the close proximity of the two curves.  This means that if we are able to influence $\bar{q}$ then it will have a corresponding effect on the CO$_2$ concentration and vice-versa.
As CO$_2$ is removed the velocity of the mixture decreases. Conservation of mass requires that the nitrogen concentration must increase (the increase is of the order of the initial concentration ratio $ {c}_{10}/ {c}_{20} \approx 1.178$). Ahead of the front the values become constant since no more CO$_2$ is removed. At the outlet we see that $ {c}_2 \approx 1.18  {c}_{20}$, $ {u} \approx 0.85  {u}_0$.

\begin{figure}
\centering
\includegraphics[width=0.7\textwidth]{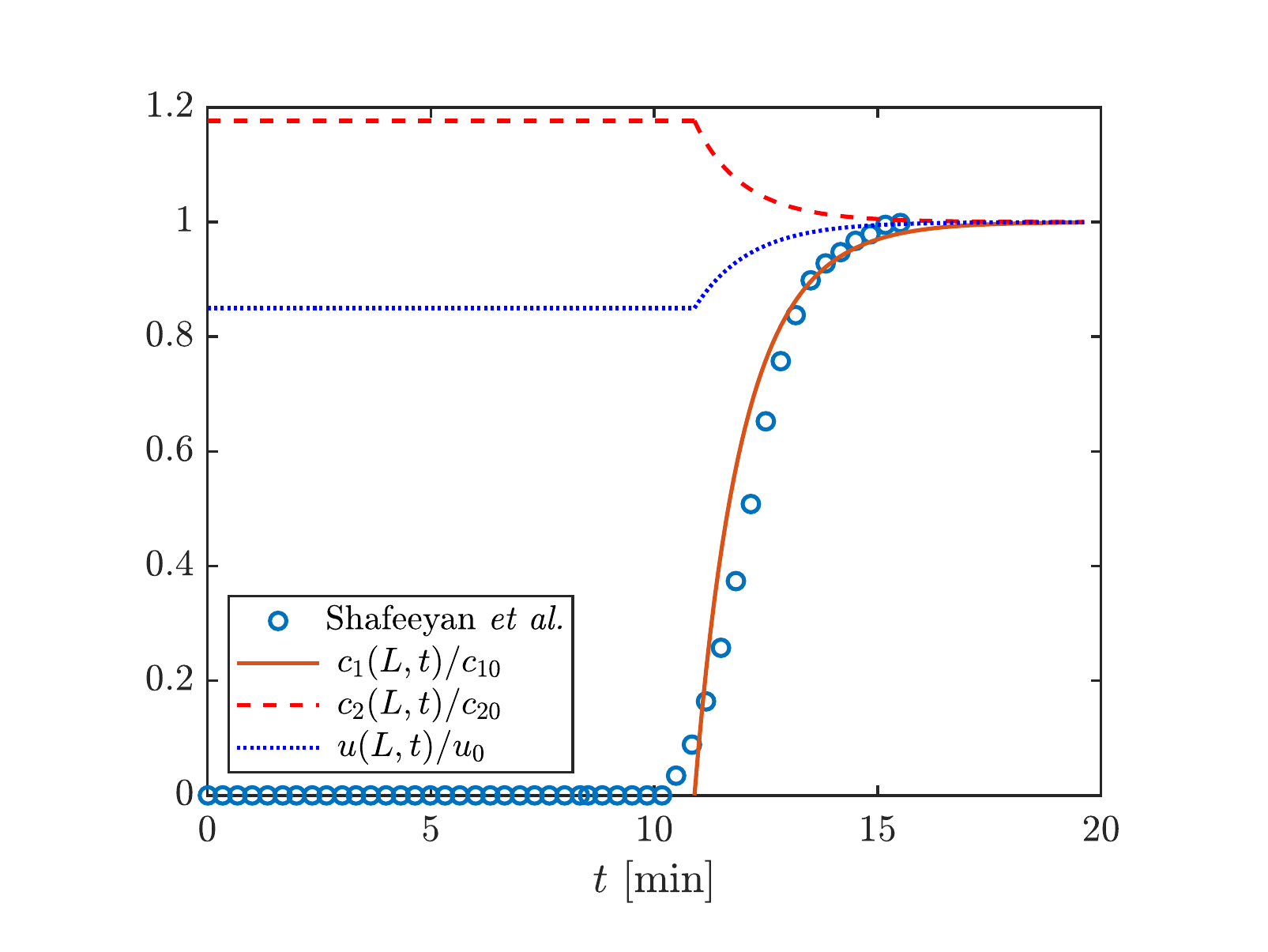}
\caption{{Concentrations $c_1(L,t), c_2(L,t)$ and velocity  $u(L,t)$ at outlet, $x=L$. All divided by their initial values. Also shown is the experimentally measured concentration of $c_1(L,t)$ (circles). }}
\label{bt}
\end{figure}

\begin{figure}
\centering
\includegraphics[width=0.7\textwidth]{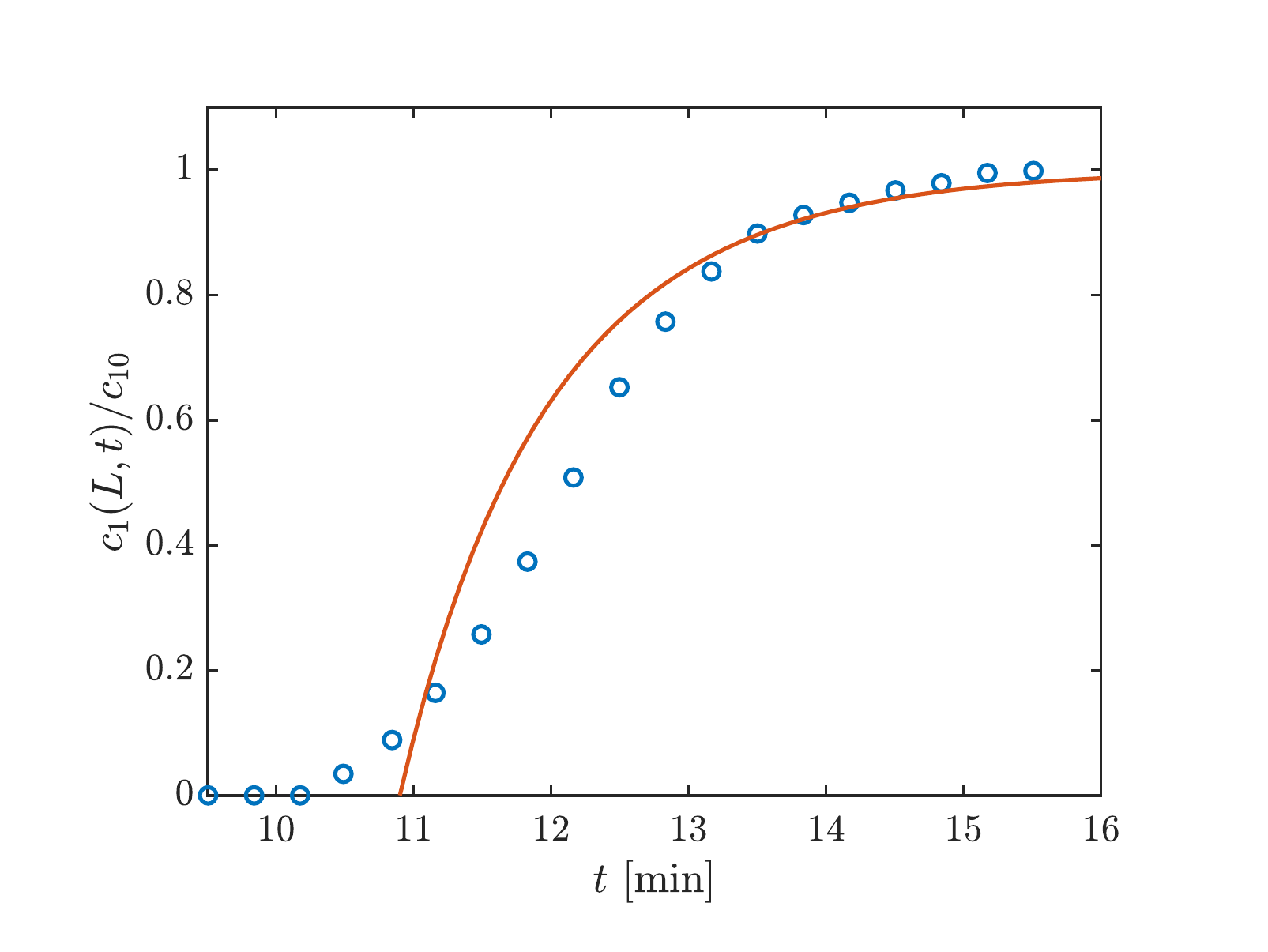}  
\caption{Blow-up of breakthrough curve, {showing the predicted ${c}_1(L,t)$ and experimental data (circles).}
}
\label{BUbt}
\end{figure}

Figure \ref{bt} shows the concentrations and velocity at the outlet. Theoretically no CO$_2$ escapes the column until $t=10.9$ minutes, after which the concentration  increases to its inlet value at around 17.5 minutes. The amount of adsorbent is not shown in the figure, since it very closely follows the CO$_2$ curve. Both $ {c}_1$ and $\bar{q}$ have their greatest rate of increase at the breakthrough time, the gradient (with respect to time) then decreases monotonically to zero as the adsorbent is used up. The theoretical velocity, $u(L,t)$, is around 0.85$u_0$ until breakthrough, when it increases to the inlet value. Similarly for the nitrogen concentration, which decreases from $1.18 c_{20}$ to its inlet value. However, we should point out that the travelling wave solution is not valid at small times, so   the curves may not be accurate near $t = 0$. The correspondence between the CO$_2$ concentration and the experimental data of \cite{Shaf15} (shown as circles) appears very good, showing a similar level of accuracy to previous published results which typically present curves over a large time range (which then makes the curves more difficult to distinguish). In Figure \ref{BUbt} we show a close-up of the comparison between the predicted CO$_2$ concentration predictions and experiment. At this scale we observe qualitative differences in the results. In particular we note that the experimental breakthrough is a more gentle process, with $ {c}_1$ at first slowly increasing to cross the theoretical prediction before increasing in gradient so the two sets of results coincide. We will discuss these differences in more detail in \S \ref{AltSec}.

\begin{figure}
\centering
\includegraphics[width=0.7\textwidth]{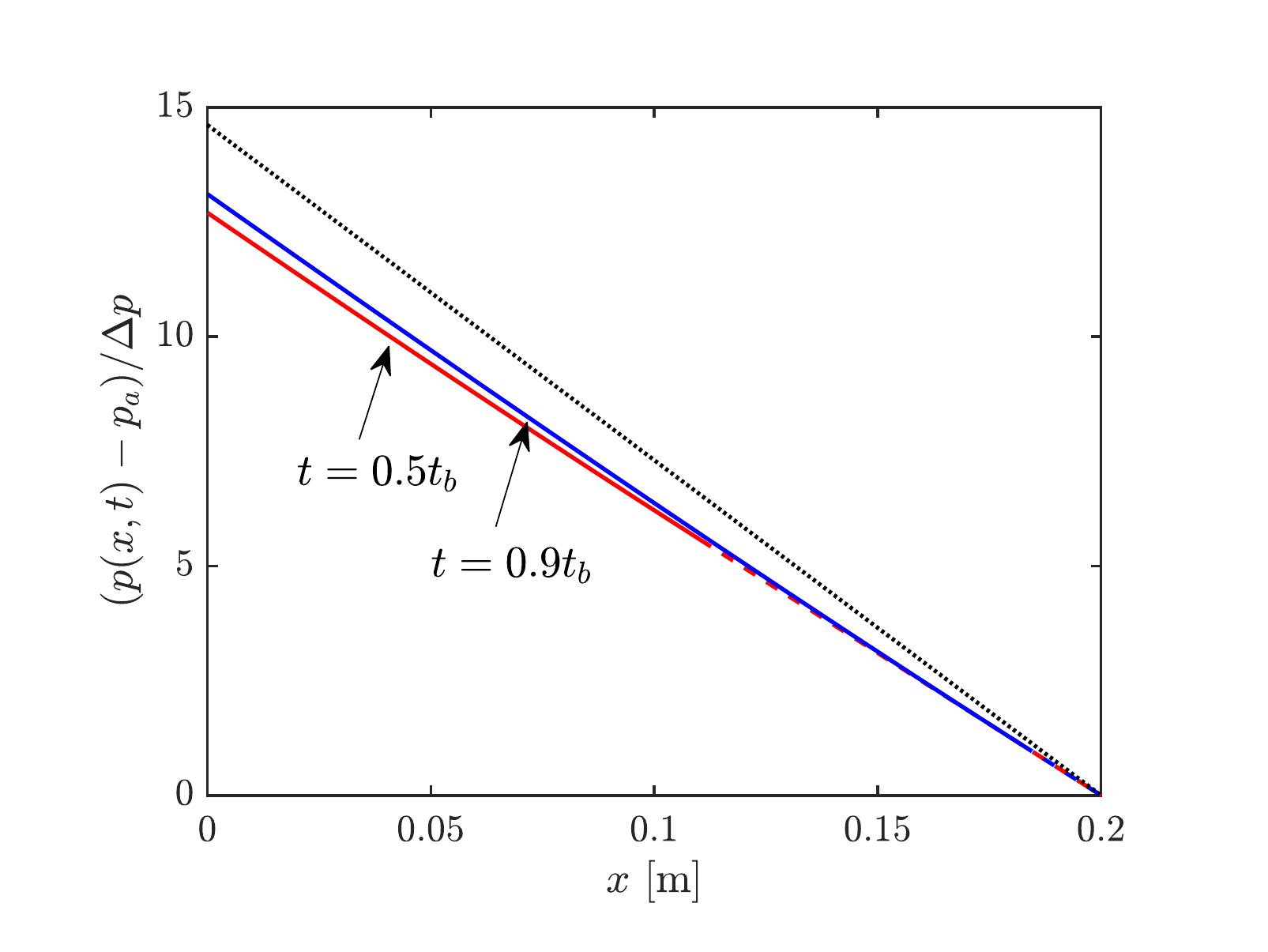}
\caption{Pressure variation along the column for $t=0.5t_b$ {(red), $t=0.9t_b$ (blue). The dashed section, near the end of the column, shows the pressure ahead of the front. The standard linear profile (dotted line) is also shown.}}
\label{PFig}
\end{figure}

In Figure \ref{PFig} we show the pressure profile at times $t=0.5 t_b, 0.9 t_b$. The solid line depicts the pressure behind the front, the dashed line that ahead of it. The dotted line is the linear profile calculated from
$- {p}_x = \beta  {u}_0$ which indicates  $ {p}= {p}_a + \beta  {u}_0 (L-x)$. This is the profile that would be observed if velocity variation were neglected, it is also the limit of the present theoretical curves for large time (when the adsorbent is full and so the velocity is constant).  The variation of $ {p}(0,t)$ verifies our earlier statement that a flow meter will have to vary the pressure to maintain a fixed flow rate.

\begin{figure}
\centering
\includegraphics[width=0.5\textwidth]{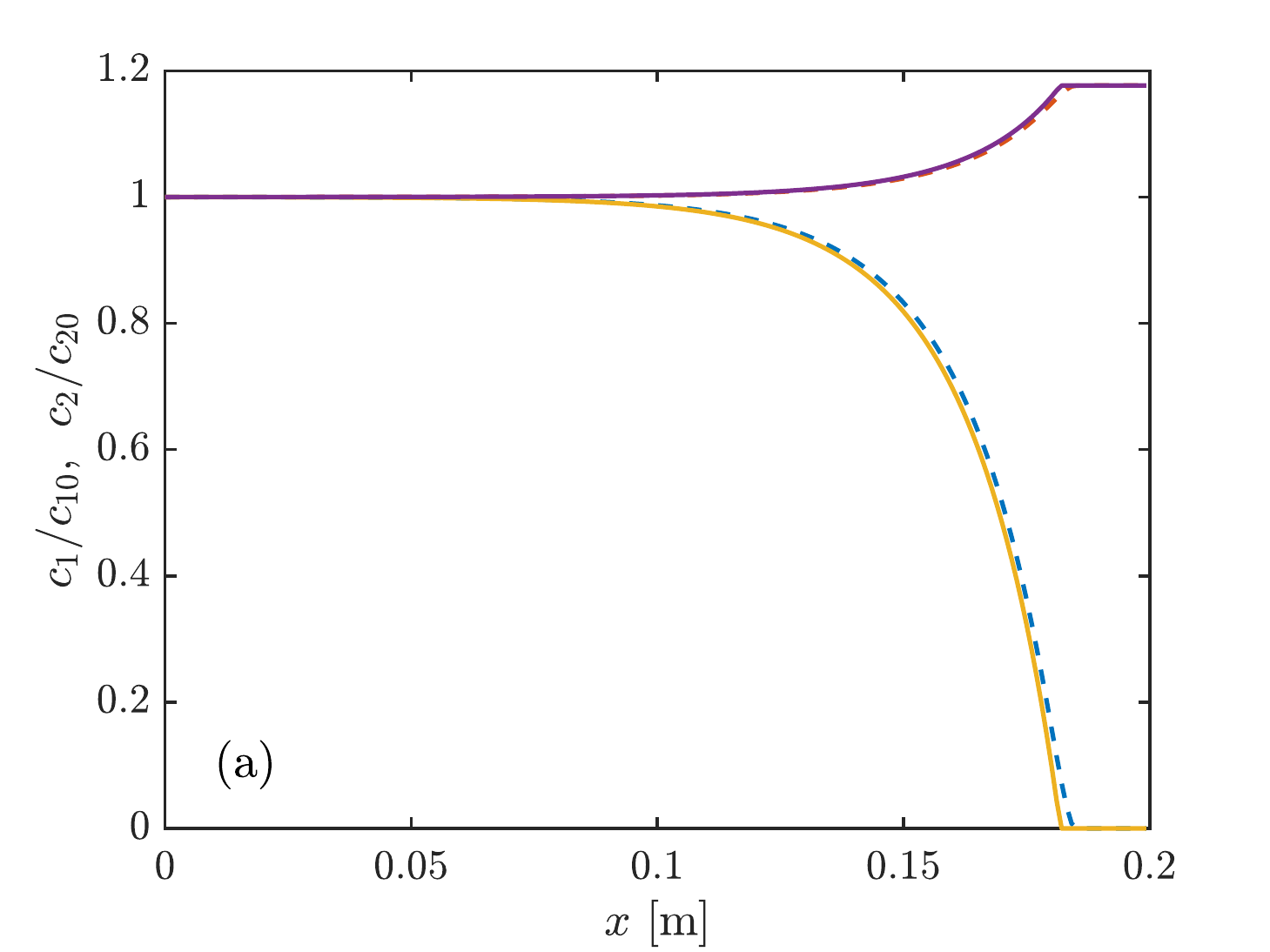}\includegraphics[width=0.5\textwidth]{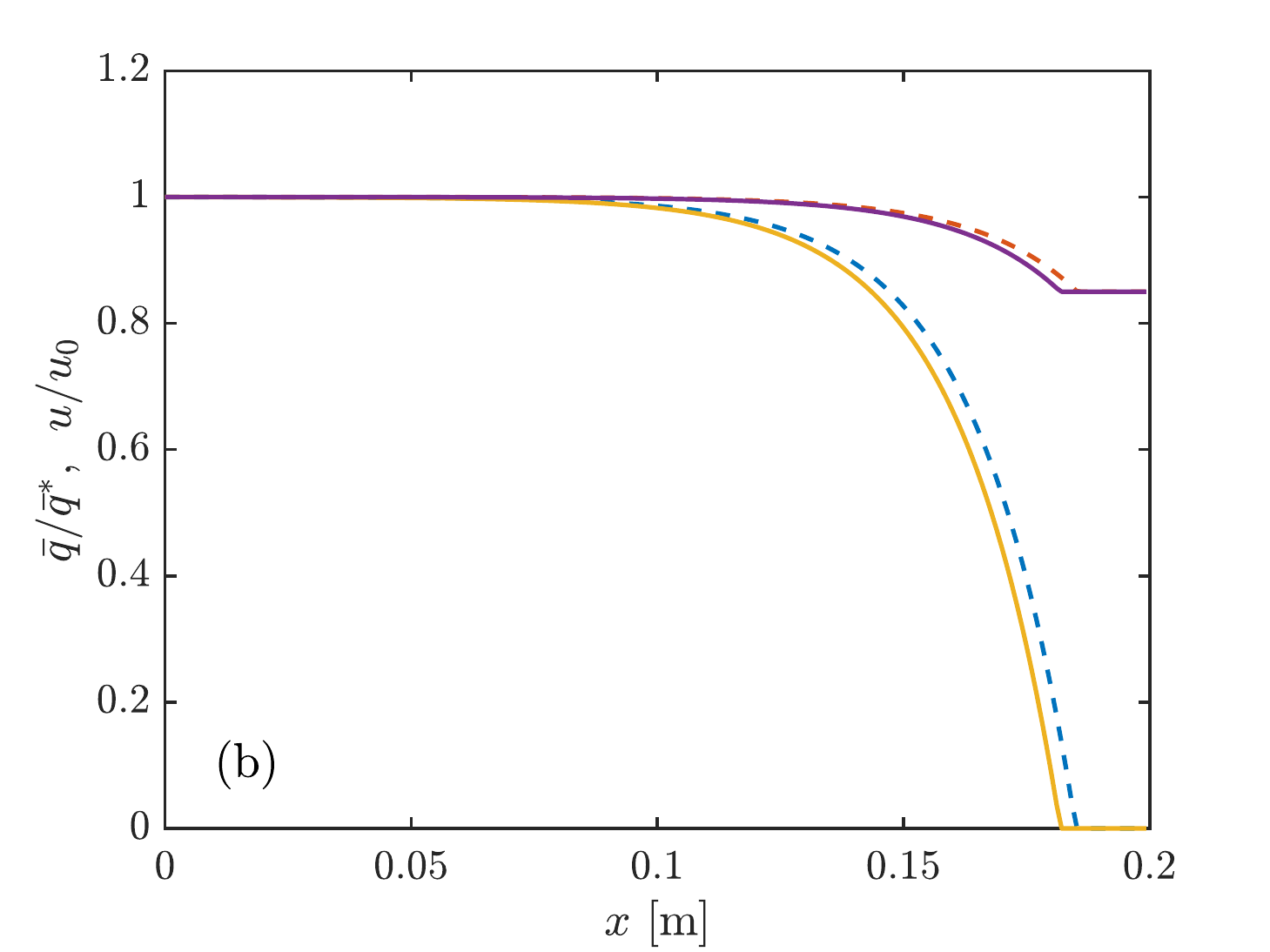}
\caption{Comparison of numerical (dashed) and travelling wave (solid) results for the same conditions as those of Figure~\ref{c12u}{: (a) shows $c_1(x, 0.9t_b), c_2(x,0.9t_b)$, (b) shows $\overline{q}(x,0.9t_b), u(x,0.9t_b)$.}}
\label{NumTW}
\end{figure}

To convince the numerically minded reader of the accuracy of the travelling wave solution, in Figure \ref{NumTW} we compare numerical and travelling wave results. The curves correspond to those presented in Figure \ref{c12u}. The dashed lines represent  the numerical solution, the solid lines the travelling wave. Full details of the system solved numerically and the scheme are provided in Appendix A. In Figure \ref{NumTW}a) we compare the CO$_2$ and N$_2$ concentrations, the CO$_2$ concentration  obviously corresponds to the curves reaching zero around 0.18m. In the first runs of the code it turned out that there was a small discrepancy: the numerical solution moved slightly  faster  than the travelling wave, by around 4\%. The main difference between the two methods comes from the retention of $\delta_1, \delta_2$ in the numerical scheme. The values $\delta_1 =0.027, \delta_2 = 0.036$ suggest errors of the order 3.6\%, so this discrepancy is entirely in line with the approximations made.
The diffusion parameter $\delta_2$ controls the spread of the front while $\delta_1 = {\cal L} k_q/u_0 =  c_{1}/((1-\epsilon) \rho_q q_0^*$ relates to the speed.  The parameter for which we have the least information is $\rho_q$ consequently we increased this by 4\% to $\rho_q = 338$kg/m$^3$. With this adjustment we achieve the excellent agreement between numerics and the analytical solution.  Figure \ref{NumTW}b) shows the variation of velocity (top curves) and available adsorbate (bottom curves). Again the agreement is excellent. Consequently we may state that the travelling wave provides solutions within the accuracy of neglected terms, where here the largest was 0.036. If even higher accuracy is required then the numerical solution may be used to fine tune the density of adsorbate.

\section{Alternative forms of mathematical model}\label{AltSec}

There exist a variety of breakthrough models designed to model different sorption processes. Typically they are based on the probability of a component escaping and the amount of material available for mass transfer. For example in \cite{Yoon} the assumptions on the probability of escape lead to
a standard logistic equation for the concentration $ {c}_{1t} = k  {c}_1 ( {c}_{10}- {c}_1)$.
In this section we discuss a number of mathematically equivalent breakthrough models and also derive the form of the present model appropriate for describing incompressible fluid flow. The models are then compared with experimental data for the adsorption of amoxicillin and a dye from water. It is shown that in these two examples the form of the previous models is incapable of capturing the whole breakthrough curve, whereas the forms (compressible and incompressible) presented in this paper both provide a good approximation.

\subsection{Previous analytical models}

An early, classic model to describe the concentration and amount absorbed was developed by Bohart and Adams \cite{Bohart}. They wrote down a constant velocity model where the time derivative and diffusion terms are neglected from the conservation equation for $c_1$ and the absorption rate depends on the amount already absorbed and the available absorbate
\bea
\pad { {c}}{ x} = - \frac{k_{BA}}{v} (\bar{q}^*-\bar{q}) ~  {c} ~ , \qquad \pad {\bar{q}}{ t} = k_{BA} (\bar{q}^*-\bar{q})~  {c} ~,
\eea
where $\bar{q}^*$ is constant.
They provide the solution
\begin{align}
{c}&= \frac{ {c}_0}{1-\exp(-k_{BA}  {c}_0 t) + \exp(k_{BA}(\bar{q}^*x/v- {c}_0 t))} ~, \label{cBAEq}\\
\bar{q}&= \frac{\bar{q}_0}{1-\exp(k_{BA}(\bar{q}^*x/v) + \exp(k_{BA}( {c}_0 t -\bar{q}^*x/v))}
\end{align}
see \cite[eq. (21,22)]{Bohart}. The breakthrough curve is obtained by setting $x=L$.
This is a much abused result and is  often misquoted, as discussed in \cite{Chu10}. In fact even in \cite{Chu10} a \lq simplified' version is studied which results from neglecting the first exponential in the denominator in \eqref{cBAEq}
\bea
{c}\approx  \frac{ {c}_0}{1 + \exp(k_{BA}(\bar{q}^*x/v- {c}_0 t))}  \label{cBAapprox} ~ .
\eea
This may be justified by assuming a sufficiently large time such  that $\exp(-k_{BA}  {c}_0 t) \ll 1$. Equation \eqref{cBAapprox} is often referred to as  the Thomas model \cite{Ahmed,Han,Smaranda}.
If we divide the top and bottom by the exponential term and again assume that an exponential term is small,
$\exp(-k_{BA}(\bar{q}^*x/v- {c}_0 t)) \ll 1$, then we obtain a formula which is only valid for small concentrations
\bea
{c}\approx   {c}_0\exp(-k_{BA}(\bar{q}^*x/v- {c}_0 t)) \label{cBAapprox2} ~
\eea
see \cite{Ahmed,Han,Smaranda} for example.
It is equivalent to the Wolborska model \cite{Patel,Smaranda}.

In arriving at \eqref{cBAapprox} we assumed $\exp(-k_{BA}  {c}_0 t) \ll 1$, usually this is justifiable after substituting for the  parameter values and considering a  sufficiently large time.
In arriving at \eqref{cBAapprox2} we assumed that a second exponential is small, leaving an expression where the concentration is proportional to this neglected exponential, hence it is only valid for small concentrations. Equation \eqref{cBAapprox2}
is, rather harshly, usually termed the Bohart-Adams equation instead of their more widely applicable result \eqref{cBAEq}. Since it only holds for small $c$ it is often stated that their model is  only valid near the start of breakthrough.

We now focus on the breakthrough curve and write equation \eqref{cBAapprox} at $x=L$ in a slightly more general form
\begin{align}
{c}_1 &=  {c}_{10} \left[ \frac{1}{1+ A_0\exp(-A_1 t)} \right]  ~ ,\label{GenForm}
\end{align}
where for the Bohart-Adams model $A_0=\exp(k_{BA}\bar{q}^*L/v), A_1 = k_{BA}  {c}_0$.
This form also covers the  Yoon-Nelson, Thomas and Bed Depth Service Time models where the parameters $A_0, A_1$ have slightly different interpretations in each case (see Table 3 in the review paper \cite{Patel}). However, since each involve some fitting to experimental data they are mathematically equivalent. Similarly we may write the present model, equation \eqref{c1Break}, as
\begin{align}
{c}_1 &=  {c}_{10} \left[ \frac{1-A_2 \exp(-A_3 t)}{1- A_4 \exp(-A_3 t)} \right]  ~ ,\label{GenForm2}
\end{align}
where
$A_2 = \exp(k_q t_b), A_3 = k_q$, $A_4 =  (R_g T  {c}_{10}/ {p}_a)A_2 $.

\subsection{Incompressible fluid or negligible adsorption}

The model derived in \S \ref{GovSec} allows for velocity variation due to the removal of significant amounts of material from the fluid, this affects the density and so is equivalent to a compressible flow. If we treat the fluid as incompressible or only consider a small amount of mass transfer then the leading order non-dimensional problem is governed by
\bea
\pad{\hat{c}_1}{x} = - \pad{\hat{q}}{\hat{t}} ~ , \qquad \pad{\hat{q}}{\hat{t}} = 1 - \hat{q} ~ .
\eea
The non-dimensional velocity $\hat{u}=1$ and the scale $ {u}_0 = Q_0/(\epsilon \pi R^2)$. The travelling wave analysis then leads to the incompressible form of equation \eqref{c1Break}
\begin{align}
{c}_1 &=  {c}_{10} \left(1- A_5\exp(-A_6 t) \right)   ~ ,\label{Genform3}
\end{align}
where $A_5 = \exp(k_q t_b), A_6 = k_q$.

\subsection{Results}

\begin{figure}
%
\includegraphics[width=0.5\textwidth]{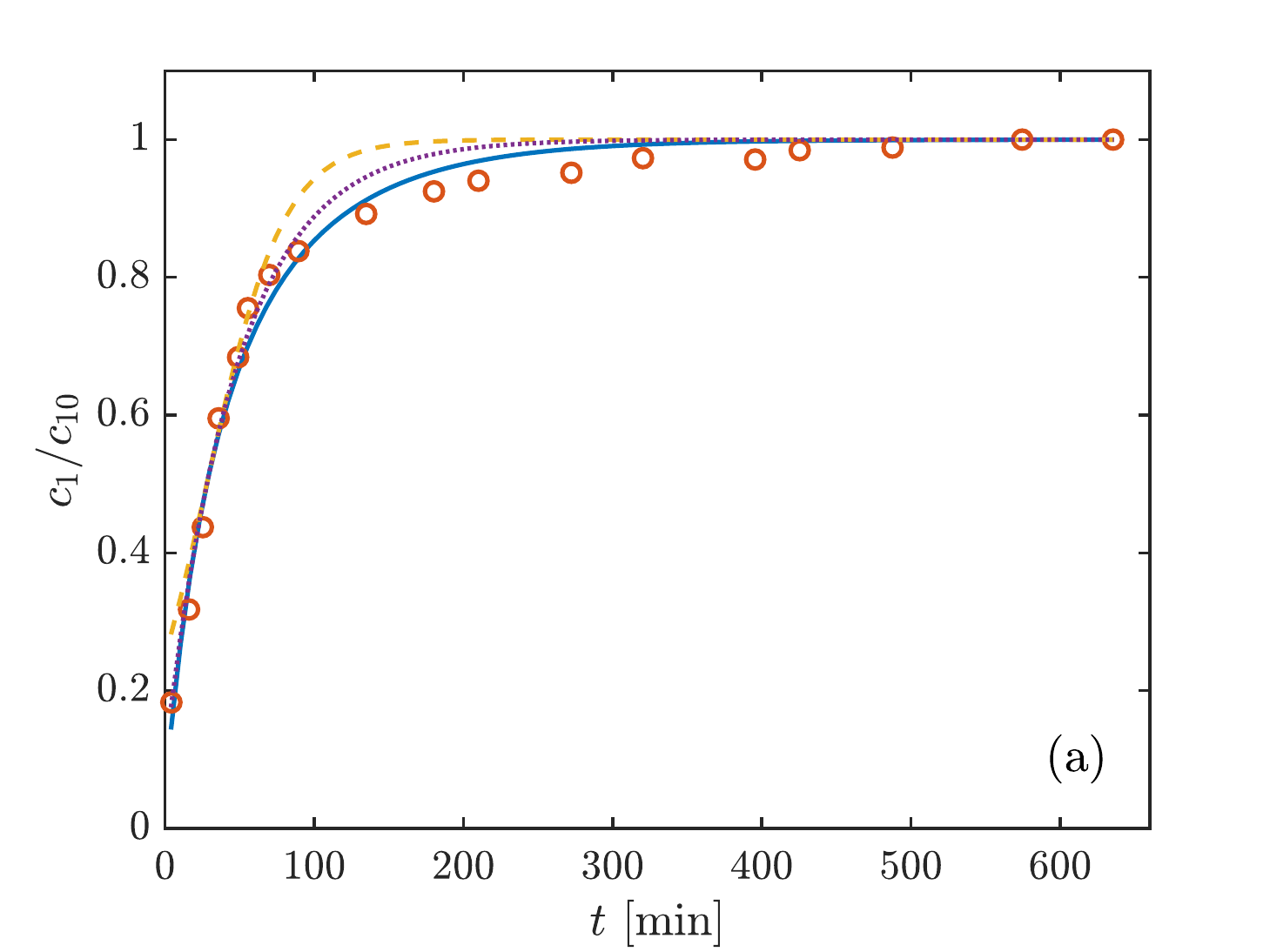}\includegraphics[width=0.5\textwidth]{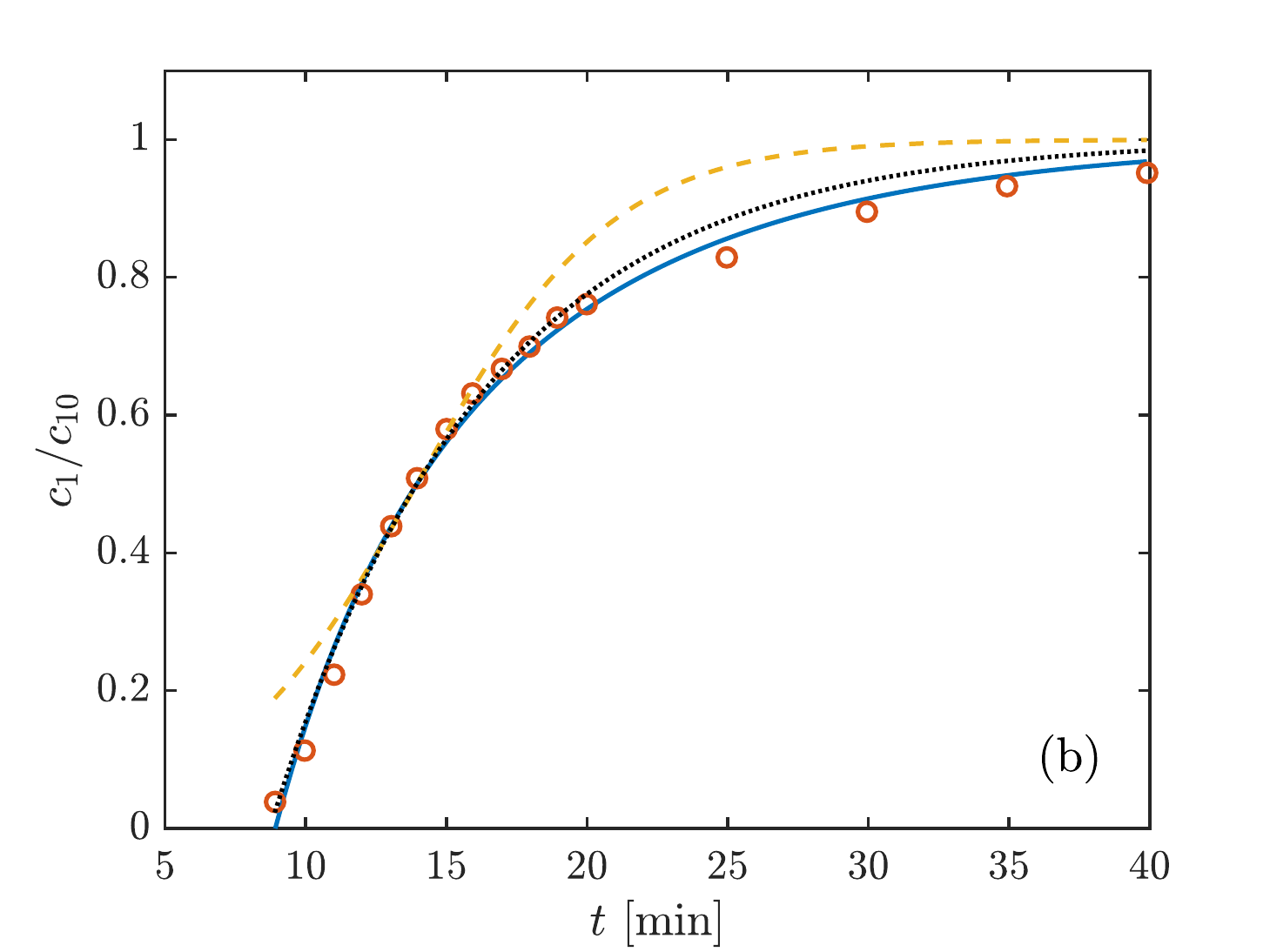}
\caption{Experimental data (circles) for the adsorption of  a) congo red dye in water by soil, \cite[Fig. 4]{Smaranda}, b) amoxicillin in water by activated carbon, \cite[Fig. 9]{Espina}. Lines are least-squares fit to equation \eqref{GenForm2} (solid), equation \eqref{GenForm} (dashed), equation \eqref{Genform3} (dotted).}
\label{fig:5}
\end{figure}

In Figures \ref{fig:5}a), b) we show comparisons of the current model against data for the removal of dye \cite{Smaranda} and amoxicillin \cite{Espina}. In each case we determine the time when the concentration is half the inlet value, $t_{1/2}$, from the experimental data (for the dye $t_{1/2}=28 \times 60$ s, for the amoxicillin  $t_{1/2}=13.95 \times 60$ s) and use this to eliminate an unknown from each model. For equation \eqref{GenForm} we find  $A_0 = \exp(A_1 t_{1/2})$, for equation \eqref{GenForm2} $A_4= 2A_2-\exp(A_3 t_{1/2})$ and for equation \eqref{Genform3} $A_5 = \exp(A_6 t_{1/2})/2$. Then we apply a least-squares fit to determine the remaining unknowns.
The experimental points in Figure \ref{fig:5}a) relate to the removal of congo red dye from solution after being passed through soil, see \cite[Fig. 4, $H=5$\,cm]{Smaranda}. The solid line comes from the current model, equation \eqref{GenForm2} with $A_2=0.98$,  $A_3 = 2.2\times 10^{-4}$s$^{-1}$, the dashed line corresponds to equation \eqref{GenForm} with $A_1=6.7 \times 10^{-4}$s$^{-1}$, the dotted line \eqref{Genform3} with $A_6 = 3.5 \times 10^{-4}$s$^{-1}$.
The experimental points in Figure \ref{fig:5}b) relate to the removal of amoxicillin from water using activated carbon, see \cite[Fig. 9]{Espina}. Again the solid line represents the current model, equation \eqref{GenForm2}, now with
$A_2=2.38$,  $A_3= 0.0016$s$^{-1}$,  the dashed line equation \eqref{GenForm} with $A_1=0.0048$s$^{-1}$, the dotted line \eqref{Genform3} with $ A_6 = 0.022$s$^{-1}$. In both graphs the best fit is provided by the current compressible flow model. Of the one parameter models the best fit is the present model for incompressible flow.

\subsection{Mass transfer models}

In the above we carried out a least-squares fit to determine the system unknowns. Other researchers use different methods, such as linear regression. Whatever the method it is clear that the form of equation \eqref{GenForm}, which describes at least four different previous models, is not capable of producing a better fit to the data used in Figure \ref{fig:5} than either of the two current models. However, this is not to say that the present model solves all problems.

Consider the form of the adsorption equation $\bar{q}_t = k_q (\bar{q}^*-\bar{q})$. The equilibrium adsorption depends on the total pressure which varies throughout the column, as shown in Figure \ref{PFig}. However, it is usually the breakthrough curve which is measured, this occurs at the column outlet where the pressure is approximately constant throughout the process. Since $\bar{q}$ increases monotonically from zero to $\bar{q}^*$ the rate $\bar{q}_t$ is greatest at first breakthrough. From the non-dimensionalisation it is clear that $ {c}_1$ differs from $\bar{q}$ by a factor of around $(1-\delta_{45})$ where $\delta_{45}$ is small. Consequently the concentration variation with time must have a similar form to that of $\bar{q}$ (this is apparent from Figure \ref{c12u}) and the highest gradient in concentration therefore also occurs at first breakthrough. This means that, although the time of first breakthrough will depend on system parameters such as the flow-rate, column length, initial concentration, void fraction etc, \emph{it is the form of the mass transfer model that primarily determines the shape of the breakthrough curve}.

The two examples shown in Figure \ref{fig:5} involve experimental data where the gradient $\partial  {c}_1/\partial t$ is greatest at first breakthrough, so the present linear driving force model results in an excellent fit. The data presented in Figure \ref{BUbt} indicates $\partial {c}_1/\partial t \approx 0$ at first breakthrough. Our model does not accurately capture this. Given that the breakthrough curve is similar in form to the mass transfer model, this behaviour will therefore be replicated with a logistic type model as suggested in \cite{Bohart}
\bea
\bar{q}_t = k  {c}_1 (\bar{q}^*-\bar{q}) ~ .
\label{qlog}
\eea
This form has a zero adsorption rate when there is zero concentration or no transfer sites available, and a maximum close to the middle of the process. So perhaps the best system to describe CO$_2$ transfer in a column will involve the current set of equations but with a mass transfer model such as equation \eqref{qlog}. Whilst the transfer of pollutants from a liquid solution best follows the present model.

\section{Conclusion}

In this paper we have developed a mathematical model to describe isothermal mass transfer from a fluid flowing through a porous  medium contained within a cylindrical tube. The model was kept relatively general, to permit the inclusion of adsorption and absorption processes and also the removal of relatively large quantities of material, such that the velocity and pressure vary nonlinearly along the column.

Since the model permits the removal of a significant amount of the fluid it is suitable to gas pollutant studies. As the amount removed becomes smaller then the model may be reduced to one more appropriate to the removal of contaminants in aqueous solutions. 
Hence we first validated it against experimental data for CO$_2$ removal by adsorption.
Subsequently we considered contaminant removal from aqueous solutions.

{Perhaps the key limitation of this work is the assumption of an isothermal reaction. In a number of studies on carbon capture it has been shown to be a small effect, similarly, with removal of trace quantities from a liquid it is clearly small however there are, no doubt, situations where this will not be an appropriate assumption. The travelling wave solution cannot capture the very early time behaviour. This may not be viewed as a problem since practical sorption equipment usually runs for very long periods and the start-up is of little interest. The model reduction was based on the size of the non-dimensional parameters, their relative size may change for different materials and experimental set-ups. Consequently they should be checked whenever a new process is investigated.} 

It was shown that sufficiently far from the inlet a travelling wave solution holds, thus there is no need for a full numerical solution. To verify this we compared the travelling wave solution against the numerics. Results showed that the difference, below 4\%, was exactly in line with the terms neglected in the analytical approximation. The numerical solution therefore turned out to be most useful for fine-tuning the value of the density of material transferred to the column or the saturation value, otherwise the travelling wave appears sufficiently accurate.

The travelling wave solution was compared against experimental data for the removal of amoxicillin and dye from water. The agreement was excellent, it also outperformed standard previous models.

A key result of the analysis is that the concentration of the material to be removed closely follows the amount of sorbate available. In the CO$_2$ example the difference was less than 10\%.
The experimentally observed breakthrough curve may then be used to guide the form of the mass transfer model. For example, in the cases of amoxicillin and dye removal the breakthrough curve had its steepest gradient at first breakthrough which then slowly decreased to zero: this suggests a kinetic relation of the form $q_t \propto q^*-q$. In the case of CO$_2$ removal published data often shows a small gradient at first breakthrough, suggesting $q_t \propto c (q^*-q)$ or $q (q^*-q)$. Since the pressure near the outlet is approximately ambient the value of $q^*$ is constant (with respect to pressure), which can simplify the breakthrough calculation  (there may still be some temperature variation, which was not considered in this paper).

The 
analytical solutions provided in our analysis, the gas concentrations, gas velocity, amount of sorbate and pressure as well as the front velocity, have been shown to accurately describe the evolution in a cylindrical column. This means that we now have explicit expressions to describe the role played by the operating parameters which may then be used to improve or optimise the process.

\section*{Acknowledgements}

F. Font acknowledges that the research leading to these results
has received funding from la Caixa Foundation. T. G. Myers acknowledges financial support from the Ministerio de Ciencia e Innovación, Spain Grant No. MTM2017-82317-P.

\appendix
\section{Numerical solution}

Numerically solving  the full model \eqref{NDqeq}-\eqref{ic_non}  can be relatively complicated,  mainly due to the nonlinearities in the momentum equation \eqref{NDpxeq}. However, a reduced version of the model can be formulated by neglecting terms of order $10^{-3}$  by setting  $\delta_3=\delta_6=\delta_8=0$ in \eqref{NDqeq}-\eqref{ic_non}. The neglect of these terms indicates errors of the order 0.1\% when compared to a solution of the full system. 

The governing equations of the reduced model are  
\begin{align}
\delta_1 \frac{\partial \hat{c}_1}{\partial\hat{t}} + \frac{\partial }{\partial \hat{x}} (\hat{u}\hat{c}_1) &= \delta_2 \frac{\partial^2 \hat{c}_1}{\partial \hat{x}^2} - \frac{\partial \hat{q}}{\partial\hat{t}}\,,\label{key1} \\ 
\delta_1 \frac{\partial \hat{c}_2}{\partial\hat{t}} + \frac{\partial }{\partial \hat{x}} (\hat{u}\hat{c}_2) &= \delta_2 \frac{\partial^2 \hat{c}_2}{\partial \hat{x}^2}\,, \label{key2}\\ 
\frac{\partial \hat{q}}{\partial\hat{t}} &= (1-\hat{q})\,, \label{key3}\\ 
1 &= \delta_4 (\hat{c}_2+\delta_5 \hat{c}_1)\,,\label{key4}\\
-\frac{\partial\hat{p}}{\partial \hat{x}} &= \hat{u}\,,\label{key5}
\end{align}
with the boundary conditions 
\begin{align}
1 &= \hat{u}|_{\hat{x}=0} \hat{c}_1|_{\hat{x}=0} - \delta_2 \left.\frac{\partial \hat{c}_1}{\partial \hat{x}}\right|_{\hat{x}=0}\,,\qquad \left.\frac{\partial \hat{c}_1}{\partial \hat{x}}\right|_{\hat{x}=l} = 0\,,\label{bc1}\\
1 &= \hat{u}|_{\hat{x}=0} \hat{c}_2|_{\hat{x}=0} - \delta_2 \left.\frac{\partial \hat{c}_2}{\partial \hat{x}}\right|_{\hat{x}=0}\,,\qquad \left.\frac{\partial \hat{c}_2}{\partial \hat{x}}\right|_{\hat{x}=l} = 0\,,\label{bc2}\\
&
\left.\frac{\partial \hat{q}}{\partial \hat{x}}\right|_{\hat{x}=0} = \left.\frac{\partial \hat{q}}{\partial \hat{x}}\right|_{\hat{x}=l} = 0\,,\label{bc3}
\end{align}
and the initial conditions given by \eqref{ic_non}. Note that \eqref{key1}-\eqref{bc3} include terms of order $10^{-2}$ previously neglected in the derivation of the travelling wave solution. Hence, the numerical solution of \eqref{key1}-\eqref{bc3} can be used to validate the accuracy of the travelling wave solution.  

Expression \eqref{key4} leads to 
\begin{equation}
\hat{c}_2 = \frac{1}{\delta_4}-\delta_5 \hat{c}_1\, .\label{key_c2}
\end{equation} 
Substituting \eqref{key_c2} in \eqref{key2} provides a relation between $\hat{u}$ and $\hat{c}_1$, which can be combined with \eqref{key1} to give
\begin{equation}
\frac{\partial \hat{u}}{\partial \hat{x}} = -\delta_4\delta_5 \frac{\partial \hat{q}}{\partial\hat{t}}\,. 
\label{key_u}
\end{equation}
This is consistent with the travelling wave solution \eqref{c2uqeq}.
In a similar fashion, by substituting \eqref{key_c2} in \eqref{bc2}, and using \eqref{bc1}, we obtain the following boundary condition for the gas velocity   
\begin{equation}\label{bc_u}
\hat{u}|_{\hat{x}=0} = (1+\delta_5)\delta_4 = 1\,,
\end{equation}
(after noting that $(1+\delta_5)\delta_4 = 1+ \delta_3 \approx 1$).
We can integrate \eqref{key_u} and apply \eqref{bc_u} to obtain 
\begin{align}
\hat{u}= 1 - \int_{0}^{\hat{x}} \delta_5\delta_4 \frac{\partial \hat{q}}{\partial\hat{t}}\,d\hat{x} \, .
\label{key_u2}
\end{align}

We note that adsorption  can only occur in the region where $\hat{c}_1$ is present. In the travelling wave solution this corresponds to the growing region $x < s(t)$. For the numerical solution, we take a different approach and define the function 
\begin{align}\label{key}
H(\hat{c}_1) = \begin{cases}
1 \qquad \text{for}\ \hat{c}_1>0\,\\
0 \qquad \text{otherwise}\,
\end{cases}
\end{align}
which we will use to enable/disable equation \eqref{key3} if a particular region within the column contains  $\hat{c}_1$ or not (see \cite{Myers19}), thereby avoiding the difficulty of dealing with a moving boundary. 

Using \eqref{key} as a multiplying factor in \eqref{key3}, and substituting \eqref{key3} in \eqref{key_u2}, the equations of the model reduce to 
\begin{align}
\delta_1 \frac{\partial \hat{c}_1}{\partial\hat{t}} + \frac{\partial }{\partial \hat{x}} (\hat{u}\hat{c}_1) &= \delta_2 \frac{\partial^2 \hat{c}_1}{\partial \hat{x}^2} - \frac{\partial \hat{q}}{\partial\hat{t}} \label{eq_c1}\\ 
\frac{\partial \hat{q}}{\partial\hat{t}} &= (1-\hat{q})H(\hat{c}_1)\,, \label{eq_q}\\
\hat{u}&= \hat{u}|_{\hat{x}=0} - \int_{0}^{\hat{x}} \delta_5\delta_4 (1-\hat{q})H(\hat{c}_1)\,d\hat{x}\label{eq_u}
\end{align}
which are subject to the boundary conditions \eqref{bc1}, \eqref{bc3} and \eqref{bc_u}. The concentration $\hat{c}_2$ is obtained via \eqref{key_c2} and the pressure $\hat{p}$ can be constructed by numerically integrating \eqref{key5} a posteriori.

The set of equations \eqref{eq_c1}-\eqref{eq_u} are solved using second-order central finite differences in space and explicit Euler in time. The boundary conditions are discretised using one-sided second-order finite differences. The nonlinear advection term in \eqref{eq_c1} is dealt with by using an upwind scheme with the $u$ profile from the previous time step. The scheme was implemented in Matlab, using the function \texttt{trapz} for the numerical integration  of \eqref{eq_u} (at each node and time step). The choice of $\Delta t$ and $\Delta \hat{x}$ is made ensuring that the stability criteria $\Delta\hat{t}\, \delta_2/(\Delta \hat{x}^2\,\delta_1)\leq 0.5$ and $\max(\hat{u})\,\Delta\hat{t}/(\Delta \hat{x}\, \delta_1) \leq 1$ are satisfied. The plots shown in Figure~\ref{NumTW} correspond to $\Delta\hat{t}=0.5\times10^{-4}$ and $\Delta \hat{x} = 0.05$.

\bibliographystyle{plain}
\bibliography{bibcarbon}

\begin{thebibliography}{10}

\bibitem{Ahmed}
M.J. Ahmed and B.H. Hameed.
\newblock Removal of emerging pharmaceutical contaminants by adsorption in a
  fixed bed column: A review.
\newblock {\em Ecotoxicology and Environmental Safety 149 (2018) 257–266},
  149:257--266, 2018.

\bibitem{Ayawei17}
N.~Ayawei, A.N. Ebelegi, and D.~Wankasi.
\newblock Modelling and interpretation of adsorption isotherms.
\newblock {\em Journal of Chemistry}, 2017.

\bibitem{BenMansour}
R.~Ben-Mansour, M.A. Habib, O.E. Bamidele, M.~Basha, N.A.A. Qasem,
  A.~Peedikakkal, T.~Laoui, and M.~Ali.
\newblock Carbon capture by physical adsorption: Materials, experimental
  investigations and numerical modeling and simulations - {A} review.
\newblock {\em Applied Energy}, 161:225 -- 255, 2016.

\bibitem{Bohart}
G.~S. Bohart and E.~Q. Adams.
\newblock Some aspects of the behaviour of charcoal with respect to chlorine.
\newblock {\em J. Am. Chem. Soc.}, 42(3):523--544, 1920.

\bibitem{Chowd13}
Z.~Z. Chowdhury, S.~M. Zain, A.~K. Rashid, R.F. Rafique, and K.~Khalid.
\newblock Breakthrough curve analysis for column dynamics sorption of mn(ii)
  ions from wastewater by using mangostana garcinia peel-based
  granular-activated carbon.
\newblock {\em Journal of Chemistry}, 2013.

\bibitem{Chu10}
K.H. Chu.
\newblock Fixed bed sorption: Setting the record straight on the
  {Bohart–Adams} and {Thomas} models.
\newblock {\em Journal of Hazardous Materials}, 177:1006--1012, 2010.

\bibitem{Espina}
Marcela Andrea~Espina de~Franco, Cassandra~Bonfante de~Carvalho,
  Mariana~Marques Bonetto, Rafael de~Pelegrini~Soares, and Liliana~Amaral
  Feris.
\newblock Removal of amoxicillin from water by adsorption onto activated carbon
  in batch process and fixed bed column: Kinetics, isotherms, experimental
  design and breakthrough curves modelling.
\newblock {\em Journal of Cleaner Production}, 161:947--956, 2017.

\bibitem{Han}
R~Han, Y.~Wang, X.~Zhao, Y.~Wang, F.~Xie, J.~Cheng, and M.~Tang.
\newblock Adsorption of methylene blue by phoenix tree leaf powder in a
  fixed-bed column: experiments and prediction of breakthrough curves.
\newblock {\em Desalination}, 245:284--297, 2009.

\bibitem{Li}
Shuangjun Li, Shuai Deng, Li~Zhao, Ruikai Zhao, Meng Lin, Yanping Du, and Yahui
  Lian.
\newblock Mathematical modeling and numerical investigation of carbon capture
  by adsorption: Literature review and case study.
\newblock {\em Applied Energy}, 221:437 -- 449, 2018.

\bibitem{Myers19}
Tim~G. Myers, Francesc Font, and Matt~G. Hennessy.
\newblock Mathematical modelling of carbon capture in a packed column by
  adsorption.
\newblock {\em Applied Energy}, 278:115565, 2020.

\bibitem{Patel}
H.~Patel.
\newblock Fixed-bed column adsorption study: a comprehensive review.
\newblock {\em Applied Water Science}, 9(45), 2016.

\bibitem{Reza09}
Fateme Rezaei and Paul Webley.
\newblock Optimum structured adsorbents for gas separation processes.
\newblock {\em Chemical Engineering Science}, 64(24):5182 -- 5191, 2009.

\bibitem{Sarker}
Ariful~Islam Sarker, Adisorn Aroonwilas, and Amornvadee Veawab.
\newblock Equilibrium and kinetic behaviour of {CO}2 adsorption onto zeolites,
  carbon molecular sieve and activated carbons.
\newblock {\em Energy Procedia}, 114:2450 -- 2459, 2017.
\newblock 13th International Conference on Greenhouse Gas Control Technologies,
  GHGT-13, 14-18 November 2016, Lausanne, Switzerland.

\bibitem{Shaf14}
Mohammad~Saleh Shafeeyan, Wan Mohd Ashri~Wan Daud, and Ahmad Shamiri.
\newblock A review of mathematical modeling of fixed-bed columns for carbon
  dioxide adsorption.
\newblock {\em Chemical Engineering Research and Design}, 92(5):961 -- 988,
  2014.

\bibitem{Shaf15}
Mohammad~Saleh Shafeeyan, Wan Mohd Ashri~Wan Daud, Ahmad Shamiri, and Nasrin
  Aghamohammadi.
\newblock Modeling of carbon dioxide adsorption onto ammonia-modified activated
  carbon: Kinetic analysis and breakthrough behavior.
\newblock {\em Energy \& Fuels}, 29(10):6565--6577, 2015.

\bibitem{Smaranda}
Camelia Smaranda, Maria-Cristina Popescu, Dumitru Bulgariu, Teodor Malut, and
  Maria Gavrilescua.
\newblock Adsorption of organic pollutants onto a romanian soil: Column
  dynamics and transport.
\newblock {\em Proc. Safety and Environmental Protection}, 108:108--120, 2017.

\bibitem{Tan12}
L.S. Tan, A.M. Shariff, K.K. Lau, and M.A. Bustam.
\newblock Factors affecting co2 absorption efficiency in packed column: A
  review.
\newblock {\em J. Ind. and Engng Chem.}, 18:1874–1883, 2012.

\bibitem{Xu13}
Zhe Xu, {Jian-Quo} Cai, and {Bing-Cai} Pan.
\newblock Mathematically modeling fixed-bed adsorption in aqueous systems.
\newblock {\em J Zhejiang Univ-Sci A (Appl Phys and Eng)}, 14(3):155--176,
  2013.

\bibitem{Yoon}
{Y.H.} Yoon and {J.H.} Nelson.
\newblock Application of gas adsorption kinetics i. a theoretical model for
  respirator cartridge service life.
\newblock {\em American Industrial Hygiene Association Journal},
  45(8):509--516, 1984.

\end{thebibliography}

\end{document}